\documentclass[amsmath,amssymb,aps,pra,preprint,groupedaddress]{revtex4-1}
\usepackage{graphicx}% Include figure files
\usepackage{dcolumn}% Align table columns on decimal point
\usepackage{bm}% bold math
\usepackage{longtable}
\usepackage{ulem}
\usepackage{color,soul}

\begin{document}
%\preprint{}

%\title{Skyrmion shape deformation due to anisotropy in DM and ferromagnetic interactions}

\title{Finsler geometry modeling and Monte Carlo study of skyrmion shape
deformation by uniaxial stress}


%\email[]{koibuchi@mech.ibaraki-ct.ac.jp}
%\thanks{}
\author{Sahbi El Hog$^{1}$}
\author{Fumitake Kato$^{2}$ }
\author{Hiroshi Koibuchi$^{3}$ }
\email[]{koi-hiro@sendai-nct.ac.jp; koibuchih@gmail.com}
\author{Hung T. Diep$^{4}$}
\email[]{diep@u-cergy.fr} 

\affiliation{
  $^{1}$Laboratoire de la Mati${\grave{e}}$re Condens${\acute{e}}$e et des Nanosciences (LMCN), Universit${\acute{e}}$ de Monastir, D${\acute{e}}$partement de Physique, Facult${\acute{e}}$ des Sciences de Monastir, Avenue de l'Environnement, 5019 Monastir, Tunisia~\\
  $^{2}$Department of Industrial Engineering, National Institute of
Technology (KOSEN), Ibaraki College, Nakane 866, Hitachinaka, Ibaraki
312-8508, Japan~\\
 $^{3}$National Institute of Technology
(KOSEN), Sendai College, 8 Nodayama, Medeshima-Shiote, Natori-shi,
Miyagi 981-1239, Japan~\\
$^{4}$Laboratoire de Physique The${\acute{o}}$rique et Mod${\acute{e}}$lisation,
University of Cergy-Pontoise, CNRS, UMR 8089 2, Avenue Adolphe Chauvin,
95302 Cergy-Pontoise Cedex, France
 }


%\date{\today}


\begin{abstract}
Skyrmions   in chiral magnetic materials are topologically stable and energetically balanced spin 
configurations appearing under the presence of ferromagnetic interaction
(FMI) and Dzyaloshinskii-Moriya interaction (DMI). Much of the current
interest has focused on the effects of magneto-elastic coupling 
on these interactions under mechanical stimuli, such as uniaxial stresses
for future applications in spintronics devices. Recent studies suggest
that skyrmion shape deformations in thin films are attributed to an
anisotropy in the coefficient of DMI, such that $D_{x}\!\not=\!D_{y}$, which makes the ratio
$\lambda/D$ anistropic, where the coefficient of FMI $\lambda$ is isotropic.
It is also possible that $\lambda_{x}\!\not=\!\lambda_{y}$ while $D$ is isotropic for $\lambda/D$ to be anisotropic.  
In this paper, we study this problem using a new modeling technique 
constructed based on Finsler geometry (FG). 
Two possible FG models are examined: In the first
(second) model, the FG modeling prescription is applied to the FMI
(DMI) Hamiltonian. We find that these two different FG models' results
are consistent with the reported experimental data for skyrmion deformation.
We also study responses of helical spin orders
 under lattice deformations corresponding
to uniaxial extension/compression and find a clear difference between
these two models in the stripe phase, elucidating which interaction
of FMI and DMI is deformed to be anisotropic by uniaxial stresses.
\end{abstract}
%\pacs{64.60.-i \sep 68.60.-p \sep 87.16.D-}
%\keywords{}

%\maketitle must follow title, authors, ab istract, \pacs, and \keywords
\maketitle

%\section{}
%\subsection{}
%\subsubsection{}

%----------------------------------------------------------
\section{Introduction}
\label{intro} 
%----------------------------------------------------------
Skyrmions are topologically stable spin configurations \cite{Skyrme-1961,Moriya-1960,Dzyalo-1964,Bogdanov-Nat2006,Bogdanov-PHYSB2005,Bogdanov-SovJETP1989}
observed in chiral magnetic materials such as FeGe, MnSi, etc. \cite{Uchida-etal-SCI2006,Yu-etal-Nature2010,Mohlbauer-etal-Science2009,Munzer-etal-PRB2010,Yu-etal-PNAS2012},
and are considered to be applicable for future spintronics devices \cite{Fert-etal-NatReview2017}. For this purpose, many experimental and theoretical studies have been conducted \cite{Buhrandt-PRB2013,Zhou-Ezawa-NatCom2014,Iwasaki-etal-NatCom2013,Romming-etal-Science2013} specifically on responses to external stimuli such as mechanical stresses \cite{Bogdanov-PRL2001,Butenko-etal-PRB2010,Chacon-etal-PRL2015,Levatic-etal-SCRep2016,Seki-etal-PRB2017,Yu-etal-PRB2015,Banerjee-etal-PRX2014,Gungordu-etal-PRB2016}. It has been demonstrated that mechanical stresses stabilize/destabilize or deform the skyrmion configuration \cite{Ritz-etal-PRB2013,Shi-Wang-PRB2018,Nii-etal-PRL2014,Nii-etal-NatCom2015,Chen-etal-SCRep2017}.

Effects of magnetostriction of chiral magnets are analytically studied  using  spin density wave by a Landau-type free energy model, in which  magneto-elastic coupling (MEC) is assumed \cite{Plumer-Walker-JPC1982,Plumer-etal-JPC1984,Kataoka-JPSJ1987}.  In a micromagnetic theory based on chiral symmetry breaking, anisotropy in the exchange coupling is assumed in addition to magnetostriction term to implement non-trivial effects on helical states and stabilize skyrmions \cite{Bogdanov-PRL2001,Butenko-etal-PRB2010}. Using such a model implementing MEC into Ginzburg-Landau free energy, Wang et al. reported simulation data for spins' responses under uniaxial stresses \cite{JWang-etal-PRB2018}, and their results accurately explain both the skyrmion deformation and alignment of helical stripes.

Among these studies, Shibata et al. reported an experimental result
of large deformation of skyrmions  by uniaxial mechanical stress, and
they concluded that the origin of this shape deformation is an anisotropy
in the coefficient $D$ of Dzyaloshinskii-Moriya interaction (DMI),
such that $D_{x}\!\not=\!D_{y}$ \cite{Shibata-etal-Natnanotech2015}.
Such an anisotropic DMI can be caused by uniaxial mechanical stresses, because the origin of DMI anisotropy is a spin-orbit coupling \cite{Fert-etal-NatReview2017}.  
It was  reported in Ref. \cite{Koretsune-etal-SCRep2015} that
this anisotropy in $D$ comes from a quantum mechanical effect of
interactions between electrons and atoms resulting from small strains.
Moreover, skyrmion deformation can also be explained by a DMI anisotropy in combination with antiferromagnetic exchange coupling \cite{Osorio-etal-PRB2017, Gao-etal-Nature2019}.

However, we have another possible scenario for skyrmion deformation; it is an anisotropy in the FMI coefficient $\lambda$ such that $\lambda_{x}\!\not=\!\lambda_{y}$. This direction-dependent $\lambda$  causes an anisotropy $\lambda/D$ even for isotropic $D$ as discussed in Ref. \cite{Shibata-etal-Natnanotech2015}, although the authors concluded that  anisotropy $\lambda/D$ comes form anisotropy in $D$. 
Such an anisotropy in $\lambda$,  the direction dependent coupling of FMI, also plays an important role in the domain wall orientation \cite{Vedmedenko-PRL2004}.

Therefore, it is interesting to study which coefficient of FMI and DMI should be anisotropic for the skyrmion deformation and stripe alignment by a new geometric modeling technique. On the stripe alignment, Dho et al. experimentally studied the magnetic microstructure of an ${\rm La_{0.7}Sr_{0.3}MnO_{3}}$ (LSMO) thin film and reported magnetic-force microscope images under tensile/compressive external forces \cite{JDho-etal-APL2003}.

In this paper, using Finsler geometry (FG) modeling, which is a mathematical
framework for describing anisotropic phenomena \cite{Takano-PRE2017,Proutorov-etal-JPC2018,Egorov-etal-PLA2021}, we study two possible models for the deformation of skyrmions and the alignment of
magnetic stripes \cite{Koibuchi-etal-JPCS2019}. In one of the models,
the FMI coefficient is deformed to be $\lambda_{x}\!\not=\!\lambda_{y}$ 
while DMI is isotropic, and in the other model, the DMI coefficient
is deformed to be $D_{x}\!\not=\!D_{y}$ while FMI is isotropic. 
Both model 1 and model 2 effectively render the ratio
$\lambda/D$ anisotropic for modulated states implying that a characteristic length scale
is also rendered to be anisotropic \cite{Butenko-etal-PRB2010}. 
Note also that the present FG prescription  cannot directly describe an anisotropic magnetization expected from MEC. In this sense,  FG models in this paper are different from both the standard Landau-type model of MEC and micromagnetic theory for thin films  studied in Ref. \cite{Plumer-Walker-JPC1982,Plumer-etal-JPC1984,Kataoka-JPSJ1987,Butenko-etal-PRB2010}, although these standard models implement MEC by an extended anisotropy of FMI in the sense that a magnetization anisotropy or higher order term of magnetization is included in addition to the exchange anisotropy. 

%----------------------------------------------------------
\section{Models}
\label{models} 
%----------------------------------------------------------
%----------------------------------------------------------
\subsection{Triangular lattices}
\label{lattice} 
%----------------------------------------------------------
%f-1
\begin{figure}[t]
\centering{}\includegraphics[width=8.5cm]{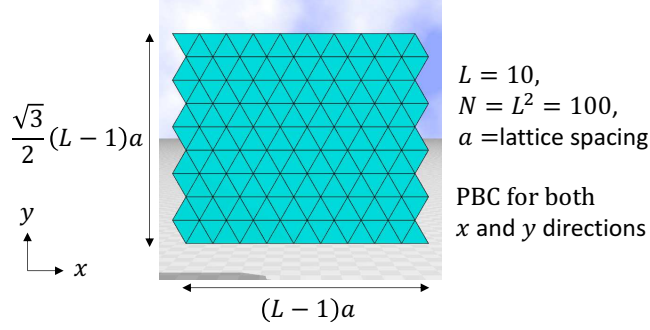}
\caption{A regular triangular lattice of size $N\!=\!L^{2}\!=\!100$, where
the total number of vertices is $L\!=\!10$ along each of the edges.
This number, $L\!=\!10$, is fixed to be very small to visualize the
lattice structure. Simulations are performed on a lattice of size
$N\!=\!10^{4}$. Periodic boundary condition (PBC) is assumed in both
directions. The lattice spacing $a$ is fixed to $a\!=\!1$ in the
simulations.}
\label{fig-1} 
\end{figure}

We use a triangular lattice composed of regular triangles of side
length $a$, called lattice spacing \cite{Creutz-txt} (Fig. \ref{fig-1}).
Triangular lattices are  used for simulating skyrmions on
thin films \cite{Okubo-etal-PRL2012,Rosales-etal-PRB2015}, where frustrated system or
antiferromagnetic interaction is assumed for studying possible mechanism of skyrmion formation on chiral magnetic materials.
However, the purpose in this paper is not the same as in  \cite{Okubo-etal-PRL2012,Rosales-etal-PRB2015}.  
On the other hand, skyrmions are known to be stabilized on thin films 
 \cite{Yu-etal-PRB2015}. On the thin film of MnSi, hexagonal skyrmion crystal is observed, 
 which can be realized on the triangular lattice. This is one of the reasons why we use triangular lattice, though
 the results in this paper are expected to remain unchanged on the regular square lattice because ferromagnetic interaction is assumed, or in other words, the system is not frustrated.  

The lattice size $N$, which is the total number of vertices, is given
by $N\!=\!L^{2}$, where $L\!-\!1$ is the total number of triangles
in both horizontal and vertical directions. The side length of the
lattice is $(L-1)a$ along the vertical direction, and $(\sqrt{3}/2)(L-1)a$
along the horizontal direction. Boundary conditions for dynamical
variables are assumed to be periodic in both directions as assumed in 3D simulations in Ref. \cite{Buhrandt-PRB2013}.  Skyrmions are topological solitons which depend on the boundary condition.  The boundary condition also strongly influences skyrmions in motion such as those in transportation. However, in our simulations, every skyrmion is only allowed to thermally fluctuate around a fixed position. For this reason,  to avoid unexpected boundary effects, we assume the periodic boundary condition. 

The lattice spacing is fixed to $a\!=\!1$ for simplicity, and the lattice size
is fixed to $N\!=\!10^{4}$ for all simulations. 
As we describe in the presentation section, the numerical results are completely independent of the lattice size up to $400\!\times\!400$ at the boundary region between skyrmion and ferromagnetic phases, and therefore, all simulations are performed on the lattice of size $100\!\times\!100$.

%----------------------------------------------------------
\subsection{The Hamiltonian and a new variable for mechanical strains}
\label{Hamiltonian-partition-function}
\begin{figure}[t]
\centering{}\includegraphics[width=8.5cm]{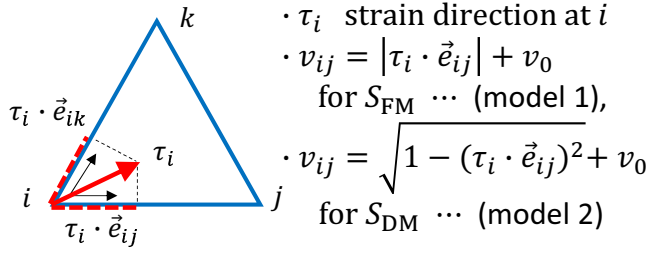}
\caption{A regular triangle of vertices $i,j,k$, and a strain direction $\tau_{i}$
at vertex $i$. The unit Finsler length $v_{ij}$ from vertices $i$
to $j$ is defined by using the tangential component $\tau_{i}\cdot{\vec{e}}_{ij}$
of $\tau_{i}$ along the direction ${\vec{e}}_{ij}$, which is the
unit tangential vector from $i$ to $j$.}
\label{fig-2} 
\end{figure}

%----------------------------------------------------------
The discrete Hamiltonian is given by the linear combination of five
terms such that 
\begin{eqnarray}
S=\lambda S_{{\rm FM}}-S_{B}+DS_{{\rm DM}}+\gamma S_{\tau}-\alpha S_{f},\quad(\alpha=1),\label{total-Hamiltonian}
\end{eqnarray}
where FMI and DMI energies $S_{{\rm FM}}$ and $S_{{\rm DM}}$ are
given in two different combinations denoted by model 1 and model 2
\cite{Koibuchi-etal-JPCS2019} (see Appendix \ref{FG-model}) 
\begin{eqnarray}
\begin{split} & S_{{\rm FM}}=\sum_{\Delta}\left[\lambda_{ij}\left(1-\sigma_{i}\cdot\sigma_{j}\right)+\lambda_{jk}\left(1-\sigma_{j}\cdot\sigma_{k}\right)+\lambda_{ki}\left(1-\sigma_{k}\cdot\sigma_{i}\right)\right],\\
 & \lambda_{ij}=\frac{1}{3}\left(\frac{v_{ij}}{v_{ik}}+\frac{v_{ji}}{v_{jk}}\right),\quad v_{ij}=|\tau_{i}\cdot{\vec{e}}_{ij}|+v_{0},\quad({\rm model\;1}),\\
 & S_{{\rm DM}}=\sum_{ij}{\vec{e}}_{ij}\cdot\sigma_{i}\times\sigma_{j},
\end{split}
\label{model-1}
\end{eqnarray}
and 
\begin{eqnarray}
\begin{split} & S_{{\rm FM}}=\sum_{ij}\left(1-\sigma_{i}\cdot\sigma_{j}\right),\\
 & S_{{\rm DM}}=\sum_{\Delta}\left[\lambda_{ij}\left({\vec{e}}_{ij}\cdot\sigma_{i}\times\sigma_{j}\right)+\lambda_{jk}\left({\vec{e}}_{jk}\cdot\sigma_{j}\times\sigma_{k}\right)+\lambda_{ki}\left({\vec{e}}_{ki}\cdot\sigma_{k}\times\sigma_{i}\right)\right],\\
 & \lambda_{ij}=\frac{1}{3}\left(\frac{v_{ij}}{v_{ik}}+\frac{v_{ji}}{v_{jk}}\right),\quad v_{ij}=\sqrt{1-\left(\tau_{i}\cdot{\vec{e}}_{ij}\right)^{2}}+v_{0},\quad({\rm model\;2}),
\end{split}
\label{model-2}
\end{eqnarray}
where FG modeling prescription is only applied to $S_{{\rm FM}}$
($S_{{\rm DM}}$) in model 1 (model 2). Note that $S_{{\rm FM}}$
in model 1 and $S_{{\rm DM}}$ in model 2 are defined by the sum over
triangles $\sum_{\Delta}$. The coefficients $\lambda$ and $D$ of
$S_{{\rm FM}}$ and $S_{{\rm DM}}$ represent the strength of FMI
and DMI. The coefficients $\lambda_{ij}$ inside the sum $\sum_{\Delta}$
of $S_{{\rm FM}}$ and $S_{{\rm DM}}$ are obtained by discretization
of the corresponding continuous Hamiltonians with Finsler metric (see
Appendix \ref{FG-model}).  $i,j,k$ of $v_{ij}$ in $\lambda_{ij}$
denote the three vertices of triangle the $\Delta$ (Fig. \ref{fig-2}).

The symbol $\sigma_{i}(\in S^{2}:{\rm unit\;sphere})$ denotes the
spin variable at lattice site $i$, which is a vertex of the triangle.
The symbol $\tau_{i}(\in S^{1}:{\rm unit\;circle})$ in $v_{ij}$
denotes a direction of strain.
 Microscopically, strains are understood
to be connected to a displacement of atoms, which  also thermally
fluctuate or vibrate without external forces. Thus, an internal variable
can be introduced to represent the direction of movement or position
deformation of atom $i$. For this reason, $\tau_{i}$ is introduced
in model 1 and model 2. A random or isotropic state of $\tau_{i}$
effectively corresponds to a zero-stress or zero-strain configuration,
while an aligned state corresponds to a uniaxially stressed or strained configuration.
The zero-strain configuration includes a random and inhomogeneous strain configuration caused by a random stress, because the mean value of
  random stress is effectively identical with zero-stress from the microscopic perspective.  
We should note that the variable $\tau_{i}$ is expected to be effective
only in a small stress region to represent strain configurations ranging
from random state to aligned state. In fact, if the variables once
align along an external force direction, which is sufficiently large,
no further change is expected in the configuration. Therefore, the
strain representation by $\tau_{i}$ is effective only in a small
stress or strain region.

One more point to note is that the variable $\tau$ is assumed to
be non-polar in the sense that it is only direction-dependent and
independent of the positive/negative direction. Indeed, the direction
of $\tau$ is intuitively considered to be related to whether the
external mechanical force is tension or compression. However, to express
an external tensile force, we need two opposite directions in general.
This assumption ($\Leftrightarrow$ $\tau$ is non-polar) is considered
sufficient because the interaction, implemented via $v_{ij}$ in Eqs.
(\ref{model-1}) and (\ref{model-2}) for $\lambda_{ij}$, is simply
dependent on $|\tau_{i}\cdot{\vec{e}}_{ij}|$ and $\left(\tau_{i}\cdot{\vec{e}}_{ij}\right)^{2}$,
respectively, where ${\vec{e}}_{ij}$ is the unit tangential vector
from vertex $i$ to vertex $j$, and hence, the interaction is dependent
only on strain directions, and independent of whether $\tau$ is polar
or non-polar.

We should note that $\lambda\lambda_{ij}$ and $D\lambda_{ij}$ in
the FMI and DMI are considered to be microscopic interaction coefficients,
which are both position ($\Leftrightarrow i$) and direction ($\Leftrightarrow ij$)
dependent. The expression of $\lambda_{ij}$ of model 1 is the same
as that of model 2, and the relation $\lambda_{ij}\!=\!\lambda_{ji}$
is automatically satisfied. However, the definitions of $v_{ij}$
are different from each other. Hence, the value of $\lambda_{ij}$
of model 1 is not always identical to that of model 2. Indeed, if
$\tau_{i}$ is almost parallel to the $x$ axis (see Fig. \ref{fig-2}),
$v_{ij}$ is relatively larger (smaller) than $v_{ik}$ and $v_{jk}$
in model 1 (model 2), and as a consequence, $\lambda_{ij}$ also becomes
relatively large (small) compared with the case where $\tau_{i}$
is perpendicular to the $x$ axis.

To discuss this point further, we introduce effective coupling constants
of DMI such that 
\begin{eqnarray}
\begin{split} & \langle D_{x}\rangle=(1/N_{B})\sum_{ij}\lambda_{ij}|\vec{e}_{ij}^{\;x}|,\\
 & \langle D_{y}\rangle=(1/N_{B})\sum_{ij}\lambda_{ij}|\vec{e}_{ij}^{\;y}|,
\end{split}
\label{anisotropy-effective-D}
\end{eqnarray}
where $\vec{e}_{ij}^{\;x}$ and $\vec{e}_{ij}^{\;y}$ are components
of $\vec{e}_{ij}=(\vec{e}_{ij}^{\;x},\vec{e}_{ij}^{\;y}){\in {\bf R}^2}$, which is the
unit tangential vector from vertex $i$ to vertex $j$ as mentioned above, and $N_{B}\!=\!\sum_{ij}1(=\!3N)$
is the total number of links or bonds. Expressions of $\langle\lambda_{x}\rangle$
and $\langle\lambda_{y}\rangle$ for FMI are exactly the same as those
of $\langle D_{x}\rangle$ and $\langle D_{y}\rangle$ in Eq. (\ref{anisotropy-effective-D}).
The symbol $\langle\cdot\rangle$ for the mean value is removed henceforth
for simplicity. Suppose the effective coupling constants $\lambda_{x}$
and $\lambda_{y}$, for $S_{{\rm FM}}$ in model 1, satisfy $\lambda_{x}>\lambda_{y}$.
In this case, the resulting spin configurations are expected to be
the same as those in model 2 under $D_{x}<D_{y}$ for $S_{{\rm DM}}$,
because an skx configuration emerges as a result of competition between
$S_{{\rm FM}}$ and $S_{{\rm DM}}$. This is an intuitive understanding
that both models are expected to have the same configuration in the
skx phase. If $\lambda_{ij}$ is isotropic or locally distributed
at random, almost independent of the direction $ij$, then the corresponding
microscopic coupling constants $\lambda_{ij}|\vec{e}_{ij}^{\;x}|$
and $\lambda_{ij}|\vec{e}_{ij}^{\;y}|$ in $D_{x}$ and $D_{y}$ of
Eq. (\ref{anisotropy-effective-D}) also become  isotropic, and consequently,
$D_{x}\!=\!D_{y}$ is expected. In contrast, if the variable $\tau$
is aligned by the external force $\vec{f}$, then $\lambda_{ij}$
becomes anisotropic or globally direction dependent, and as a consequence,
$D_{x}$ and $D_{y}$ become anisotropic such that $D_{x}\!\not=\!D_{y}$.

% f-2

We should comment that our models include a shear component of stress-effect on the coefficient $\lambda_{ij}$ in Eqs. (\ref{model-1}), (\ref{model-2}).  
To simplify arguments, we tentatively assume $v_0\!=\!0$ in model 1. Let $\vec{f}$ be $\vec{f}\!=\!(f,0)$ or parallel to ${\vec e}_{ij}$, which represents the first local coordinates axis (Fig. \ref{fig-2}), implying that $\tau_i$ is almost parallel to ${\vec e}_{ij}$ for sufficiently large $f$. Then, we have $v_{ij}\!\simeq\!|\tau_i|\!=\!1$, which represents an effect of the tensile stress $\vec{f}$ along ${\vec e}_{ij}$. 
For the same  $\tau_i$, we have $v_{ik}\!\simeq\!0.5|\tau_i|\!=\!0.5$ along ${\vec e}_{ik}$, which represents the second local coordinates axis. 
Thus, we obtain the ratio $v_{ij}/v_{ik}\!\simeq\!2$, and by moving the local coordinate origin to vertex $j$ and from the same calculation we obtain  $v_{jk}/v_{ji}\!\simeq\!2$, and therefore, $\lambda_{ij}\!=\!4/3$. Since the variables $\tau$ at all other vertices are naturally considered to be parallel to ${\vec e}_{ij}$, we have $\lambda_{ik}\!=\!1/3$ from the same argument. The fact that $\lambda_{ik}$ is non-zero under $\vec{f}\!=\!(f,0)$ is considered to be an effect of shear stress.

The other terms $S_{B}$, $S_{\tau}$ and $S_{F}$ in $S$ of Eq.
(\ref{total-Hamiltonian}) are common to both models and are given
by 
\begin{eqnarray}
\begin{split} & S_{B}=\sum_{i}\sigma_{i}\cdot\vec{B},\quad\vec{B}=(0,0,B),\\
 & S_{\tau}=\frac{1}{2}\sum_{ij}\left(1-3(\tau_{i}\cdot\tau_{j})^{2}\right),\quad S_{f}=\sum_{i}\left(\tau_{i}\cdot\vec{f}\right)^{2},\quad{\vec{f}}=(f_{x},f_{y}),
\end{split}
\label{other-Hamiltonian}
\end{eqnarray}
where $S_{B}$ is the Zeeman energy with magnetic field $\vec{B}$,
and $S_{\tau}$ is a Lebwohl-Lasher type potential \cite{Leb-Lash-PRA1972},
which is always assumed for models of liquid crystals \cite{Proutorov-etal-JPC2018}.

In $S_{f}$, $\vec{f}\!=\!(f_{x},f_{y})$ represents an external mechanical
force, which aligns the strain direction $\tau$ along the direction
of $\vec{f}$. The reason why $S_{f}$ is not linear concerning $\vec{f}$
(or $\tau$) is that the force $\vec{f}$ has a non-polar interaction
given by $S_{\tau}$. Therefore, it is natural to assume the square
type potential. In liquid crystals, such a square type potential is
also assumed for external electric fields \cite{Proutorov-etal-JPC2018}.
The coefficient $\alpha$ of $S_{f}$ in Eq. (\ref{total-Hamiltonian})
is fixed to $\alpha=1$ for simplicity. This is always possible by
re-scaling $f$ to $\sqrt{\alpha}f$.

\begin{figure}[t]
\centering{}\includegraphics[width=8.5cm]{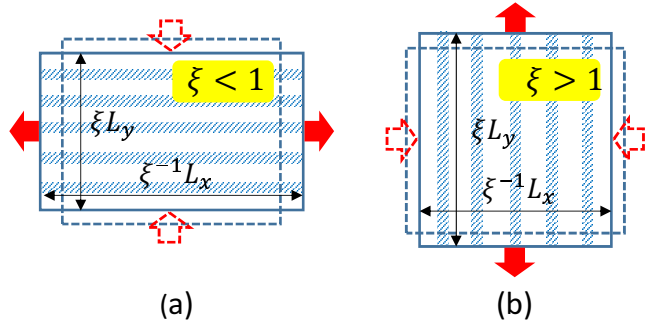}
\caption{Lattice deformations represented by (a) $\xi\!<\!1$ and (b) $\xi\!>\!1$
in Eq. (\ref{strain-deformation}). In (a) for $\xi\!<\!1$ and (b)
for $\xi\!>\!1$, the corresponding external tensile forces' direction
is horizontal and vertical, respectively. The dashed arrows represent
the direction of forces, implying that the force is assumed compressive,
and the shaded thick lines denote the stripe directions experimentally
observed and reported in Ref. \cite{JDho-etal-APL2003}. }
\label{fig-3} 
\end{figure}

Alignment of the direction of $\tau$ is essential for modeling stress-effect
in model 1 and model 2. In this paper, we assume the following two
different sources for this alignment: 
\begin{enumerate}
\item[(i)] Uniaxial stresses by ${\vec{f}}=(f,0)$ and ${\vec{f}}=(0,f)$ with
$\gamma\!=\!0\quad$ (for skyrmion deformation), 
\item[(ii)] Uniaxial strains by lattice deformation by $\xi$ with $\gamma\!>\!0\quad$
(for stripe deformation), 
\end{enumerate}
where $\xi$ in (ii) is defined by the deformations of side lengths
such that (Fig. \ref{fig-3}) 
\begin{eqnarray}
L_{x}\to\xi^{-1}L_{x},\quad L_{y}\to\xi L_{y},\label{strain-deformation}
\end{eqnarray}
where $f\!>\!0$ is assumed, implying that $\vec{f}$ is tensile,
and $L_{x}$ and $L_{y}$ are actually given by $L_{x}\!=\!(L\!-\!1)a$
and $L_{y}\!=\!(\sqrt{3}/2)(L\!-\!1)a$ as shown in Fig. \ref{fig-1}.
In both cases (i) and (ii), the variable $\tau$ is expected to be
aligned, and this alignment causes deformations in the interactions
of $S_{{\rm FM}}$ and $S_{{\rm DM}}$ to be direction dependent like
in the forms $\lambda\lambda_{ij}$ and $D\lambda_{ij}$ as mentioned
above. In the case of (i), the lattice is undeformed, implying that
$\xi$ is fixed to $\xi\!=\!1$. In this case (i), uniaxial stresses
by the external force are only applied to check the skyrmion shape
deformation, and the coupling constant $\gamma$ of $S_{\tau}$ is
assumed to be $\gamma\!=\!0$. On the contrary, in the case of (ii),
the external force $\vec{f}$ is assumed to be ineffective and fixed
to ${\vec{f}}\!=\!(0,0)$, while the parameter $\gamma$ for $S_{\tau}$
is fixed to a non-negative constant $\gamma\!>\!0$ so that $\tau$
can spontaneously align to a certain direction associated with the
lattice deformation by $\xi$. In this case (ii), $S_{{\rm DM}}$
is expected to play a non-trivial role in both model 1 and model 2,
because lattice deformations originally influence DMI. This will be
a check on whether or not a coupling of strain and spins (or magnetization)
is effectively implemented in DMI. It is clear that $S_{{\rm FM}}$
of model 2 in Eq. (\ref{model-2}) is completely independent of the
lattice deformation by $\xi$.

%f-3

The partition function is defined by 
\begin{eqnarray}
Z=\sum_{\sigma}\sum_{\tau}\exp\left[-S(\sigma,\tau)/T\right],\label{part-func}
\end{eqnarray}
where $\sum_{\sigma}$ and $\sum_{\tau}$ denote the sum over all
possible configurations of $\sigma$ and $\tau$, and $T$ is the
temperature. Note that the Boltzmann constant $k_{B}$ is assumed
to be $k_{B}\!=\!1$.

Here, we show the input parameters for simulations in Table \ref{table-1}.
\noindent
\begin{longtable}{ccp{75mm}}
\caption{List of symbols and descriptions of the input parameters.  }
  \label{table-1} \\
%------ top of the first page ----
  \hline
Symbol   & & Description  \\ \hline
\endfirsthead
%------ top of the following pages ----
\multicolumn{3}{r}{continue} \\ \hline
Symbol  & & Description (assumed typical value)  \\ \hline
\endhead
%----- botomn of the pages --------
  \hline
\multicolumn{3}{r}{continue} \\
\endfoot
%----- botomn of the final page --------
  \hline
\multicolumn{3}{r}{} \\
\endlastfoot
$T$  &  & Temperature   \\
$\lambda$ & &   Ferromagnetic interaction coefficient    \\
$D$ & &   Dzaloshinskii-Moriya interaction coefficient    \\
$B$ & &   Magnetic filed    \\
$\gamma$ & &  Interaction coefficient of $S_\tau$   \\
$f$ & &  Strength of mechanical force ${\vec f}=(f,0)$ or ${\vec f}=(0,f)$ with $f\!>\!0$ \\
$v_0$ & &   Strength of anisotropy    \\
$\xi$ & &   Deformation parameter for the side lengths of lattice: $\xi\!=\!1 \Leftrightarrow$ non-deformed    \\
\end{longtable}

\noindent
%----------------------------------------------------------
\subsection{Monte Carlo technique and snapshots\label{MC-technique}}
%----------------------------------------------------------
The standard Metropolis Monte Carlo (MC) technique is used to update
the variables $\sigma$ and $\tau$ \cite{Metropolis-JCP-1953,Landau-PRB1976}.
For the update of $\sigma$, a new variable $\sigma_{i}^{\prime}$
at vertex $i$ is randomly generated on the unit sphere $S^{2}$ independent
of the old $\sigma_{i}$, and therefore, the rate of acceptance is
not controllable. The variable $\tau$ is updated on the unit circle
$S^{1}$ by almost the same procedure as that of $\sigma$.

The initial configuration of spins is generated by searching the ground
state (see Ref. \cite{Hog-etal-JMagMat2018}). One MC sweep (MCS)
consists of $N$ consecutive updates of $\sigma$ and that of $\tau$.
In almost all simulations, $2\times10^{8}$ MCSs are performed. At
the phase boundary between the skyrmion and ferromagnetic phases,
the convergence is relatively slow, and therefore $5\times10^{8}$
MCSs or more, up to $1.6\times10^{9}$ MCSs, are performed. In contrast,
a relatively small number of MCSs are performed in the ferromagnetic
phase at large $|B|$ or high $T$ region.

%f-4
\begin{figure}[h]
\centering{}\includegraphics[width=10.5cm]{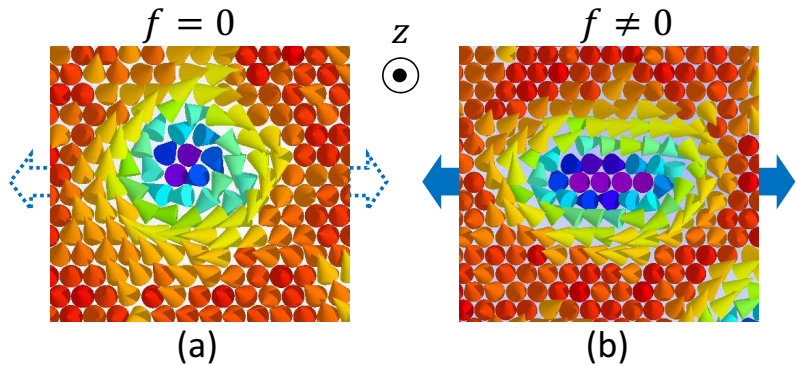}
\caption{Snapshot of skyrmions for (a)  $f\!=\!0$ and  (b) $f\!\not=\!0(=\!1.7)$ with $T\!=\!0.2$,  $D\!=\!0.45$, $\gamma\!=\!0$,  $v_0\!=\!0.7$, and $\xi\!=\!1$.  These snapshots are obtained by model 2 and the same as those obtained by model 1..  This vortex-like skyrmion is called Bloch type, which is studied in this paper.
\label{fig-4} 
}
\end{figure}
Here, we show snapshots of skyrmion configuration obtained by model 2 for $f\!=\!0$ and  $f\!\not=\!0(=\!1.7)$ in Figs. \ref{fig-4}(a),(b). The assumed parameters other than $f$ are $T\!=\!0.2$,  $D\!=\!0.45$, $\gamma\!=\!0$,  $v_0\!=\!0.7$, and $\xi\!=\!1$ for both (a) and (b). The cones represent spins $\sigma_i$, and the colors of cones correspond to $z$-component $\sigma_i^z$. We find from both snapshots that the direction of cones in the central region of skyrmions is $-z$ while it is $+z$ outside.   Skyrmion configurations of model 1 are the same as these snapshots. This vortex-like configuration (Figs. \ref{fig-4}(a)) is called Bloch type and symmetric under rotation along $z$ axis \cite{Leonov-etal-NJO2016}. In this paper, we study skyrmions of Bloch type.  

%----------------------------------------------------------
\section{Simulation results}
\label{results}
%----------------------------------------------------------
%----------------------------------------------------------
\subsection{Responses to uniaxial stress \label{uniaxial-stress}}
%----------------------------------------------------------
%----------------------------------------------------------
\subsubsection{Magnetic filed vs. Temperature diagram}
\label{BT-diagram}
%----------------------------------------------------------

\begin{figure}[h]
\centering{}\includegraphics[width=11.5cm]{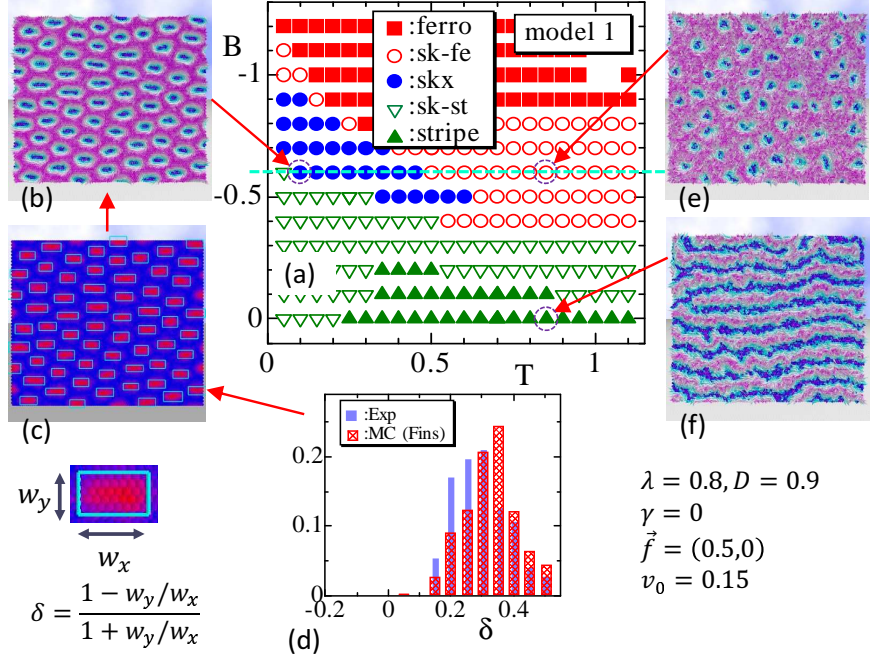}
\caption{(a)  Phase diagram of magnetic field $B$ and temperature $T$ of
model 1, (b)  snapshot obtained at $(B,T)\!=\!(-0.6,0.1)$, (c)  corresponding
snapshot to measure the shape anisotropy $\delta$ of skyrmion, (d)
histograms of $\delta$, where a reported histogram (Exp) for experimental
result in Ref. \cite{Shibata-etal-Natnanotech2015} is also plotted,
(e)  snapshot at $(B,T)\!=\!(-0.6,0.85)$, and (f)  snapshot in the
stripe phase at $(B,T)\!=\!(0,0.85)$. The symbols (skx), (str), and
(ferro) denote skyrmion, stripe, and ferromagnetic phases, respectively.
The symbols (sk-fe) and (sk-st) denote intermediate phases of skyrmion
ferromagnetic and skyrmion stripe, respectively. On the dashed horizontal
line, physical quantities are calculated in the following subsection. }
\label{fig-5} 
\end{figure}

 A phase diagram of model 1 is shown in Fig. \ref{fig-5}(a), where
the temperature $T$ and magnetic field $B$ are varied. The symbols
(skx), (str), and (ferro) denote the skyrmion, stripe, and ferromagnetic
phases, respectively. The stripe phase is the same as the so-called
helical phase, where the spins are rotating along the axis perpendicular
to the stripe direction. Between these two different phases, intermediate
phases appear, denoted by the skyrmion ferromagnetic (sk-fe) and skyrmion
stripe (sk-st) phases. The parameters $\lambda,D,\gamma,f,v_{0}$
are fixed to $(\lambda,D,\gamma,f,v_{0})\!=\!(0.8,0.9,0,0.5,0.15)$
in Fig. \ref{fig-5}(a). The applied mechanical stress is given by
$\vec{f}\!=\!(0.5,0)$, which implies that a thin film is expanded
in $x$ direction by a tensile force $f\!=\!0.5$.

The phase diagram in Fig.  \ref{fig-5}(a) is only rough estimates for identifying the regions of different states. These boundaries are determined by viewing their snapshots. For example, if a skyrmion is observed in the final  ferromagnetic configuration of simulation at the boundary region between the skx and ferro phases, this state is written as sk-fe. If two skymion states are connected to be oblong shape and all others are isolated in a snapshot, then this state is written as sk-st.  Thus, the phase boundaries in these digital phase diagrams are not determined by the standard technique such as the finite scaling analyses \cite{Janoschek-etal-PRB2013,Hog-etal-JMagMat2018}, and therefore, the order of transition between two different states is not specified.

Figure \ref{fig-5}(b) shows a snapshot of deformed skyrmions of model
1 at a relatively low temperature $T\!=\!0.1$. To measure the shape
anisotropy, we draw rectangles enclosing skyrmions, as shown in Fig.
\ref{fig-5}(c), where the edge lines are drawn parallel to the $x$
and $y$ directions. The details of how  the edge lines are drawn
can be found in Appendix \ref{skx-graphical}. This technique can
also be used to count the total number of skyrmions, at least in the
skx phase, which will be presented below. Figure \ref{fig-5}(d) shows
the distribution of shape anisotropy $\delta$ defined by 
\begin{eqnarray}
\delta=\left(1-w_{y}/w_{x}\right)/\left(1+w_{y}/w_{x}\right),\label{shape-anisotropy}
\end{eqnarray}
where $w_{x}$ and $w_{y}$ are the edge lengths of the rectangle
\cite{Shibata-etal-Natnanotech2015}. The solid histogram is the experimental
data (Exp) reported in Ref. \cite{Shibata-etal-Natnanotech2015}.
In this Ref. \cite{Shibata-etal-Natnanotech2015}, simulations were
also performed by assuming that the DMI coefficients $D$ are direction-dependent
such that $D_{x}/D_{y}\!=\!0.8$, and almost the same result with
Exp was obtained. The result of model 1 in this paper, shown in the
shaded histogram, is almost identical to that of Exp. In these histograms,
the height is normalized such that the total height remains the same.
Another snapshot obtained at higher temperature $T\!=\!0.8$ is shown
in Fig. \ref{fig-5}(e), where the shape of the skyrmion is not smooth
and almost randomly fluctuating around the circular shape. Therefore,
this configuration is grouped into the sk-fe phase, even though such
fluctuating skyrmions are numerically stable, implying that the total
number of skyrmions remains constant for long-term simulations. Figure
\ref{fig-5}(f) shows a snapshot obtained in the stripe phase. The
direction of the stripes is parallel to the direction of the tensile
force $\vec{f}\!=\!(f,0)$.

\begin{figure}[h]
\centering{}\includegraphics[width=11.5cm]{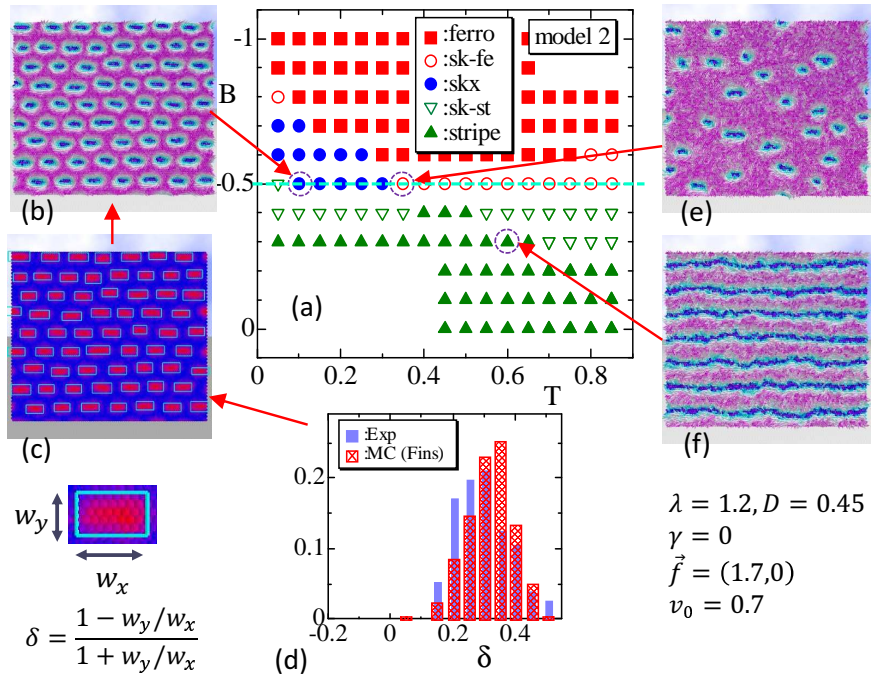}
\caption{(a)  Phase diagram of magnetic field $B$ and temperature $T$ of
model 2, (b)  snapshot obtained at $(B,T)\!=\!(-0.5,0.1)$, (c)  corresponding
snapshot to measure the shape anisotropy of the skyrmion, (d) the
corresponding histogram of $\delta$, where a reported histogram (Exp)
for experimental result in Ref. \cite{Shibata-etal-Natnanotech2015}
is also plotted, (e)  snapshot at $(B,T)\!=\!(-0.5,0.35)$, and (f)
 snapshot in the stripe phase at $(B,T)\!=\!(-0.3,0.6)$. On the dashed
horizontal line, physical quantities are calculated in the following
subsection. }
\label{fig-6} 
\end{figure}

The results of model 2 in Figs. \ref{fig-6}(a)--\ref{fig-6}(f)
are almost identical to those in Fig. \ref{fig-5}. The parameters
$\lambda,D,\gamma,f,v_{0}$ are fixed to $(\lambda,D,\gamma,f,v_{0})\!=\!(1.2,0.9,0,1.7,0.7)$
in Fig. \ref{fig-6}(a) for model 2. The unit of $T$ depends on the
ratio of $T$ and the coefficients of Hamiltonians $S_{{\rm FM}}$,
$S_{B}$, $S_{{\rm DM}}$, $S_{\tau}$ and $S_{f}$. However, the
ratios themselves cannot be compared with each other because the first
two parameters, $(\lambda,D)$ at least for model 1, are not proportional
to these parameters for model 2. In fact, $D$ in model 2 is effectively
deformed to be direction-dependent such that $DD_{x}$ and $DD_{y}$
by $D_{x}$ and $D_{y}$ in Eq. (\ref{anisotropy-effective-D}), while
$D$ in model 1 remains unchanged. Therefore, the unit of horizontal
$T$ axis in Fig. \ref{fig-5}(a) is not exactly identical but almost
comparable to that of model 1 in Fig. \ref{fig-6}(a).

The parameter $v_{0}\!=\!0.7$ assumed in $v_{ij}$ of Eq. (\ref{model-2})
for model 2 is relatively larger than $v_{0}\!=\!0.15$ in $v_{ij}$
of Eq. (\ref{model-1}) for model 1. If $v_{0}$ in model 2 is fixed
to be much smaller such as $v_{0}\!=\!0.15$ just like in model 1,
then the shape of the skyrmions becomes unstable. This fact implies
that the anisotropy of DMI caused by the FG model prescription is
too strong for such a small $v_{0}$ in model 2. Conversely, if $v_{0}$
in model 1 is fixed to be much larger, such as $v_{0}\!=\!0.7$, then
the skyrmion shape deformation is too small, implying that anisotropy
of FMI caused by the FG model prescription is too weak for $v_{0}\!=\!0.7$.

Here, we should note that the skx region in the $BT$ diagrams of Figs. \ref{fig-5} and \ref{fig-6} changes with varying $B$ at relatively low $T$ region. Indeed, if $|B|$ is increased from $B\!=\!0$ at $T\!=\!0.1$ in Fig. \ref{fig-6} for example, the connected stripes  like in Fig. \ref{fig-6}(f) start to break, and the stripe phase changes to the sk-st at $|B|\!=\!0.4$,  and the skx emerges at $|B|\!=\!0.5$  as shown in Fig. \ref{fig-6}(b). The skyrmion shape in the skx phase is oblong in $(1,0)$ direction, which is the same as the stripe direction for smaller $B$ region. This shape anisotropy of skyrmions as well as the size itself becomes smaller and smaller with increasing $|B|$, and for sufficiently large $|B|$ such as $|B|\!=\!0.8$, the skx turns to be  ferromagnetic.

%----------------------------------------------------------
\subsubsection{Temperature dependence of physical quantities}
\label{T-dependence}
%----------------------------------------------------------
%f-7
\begin{figure}[h]
\centering{}\includegraphics[width=6.5cm]{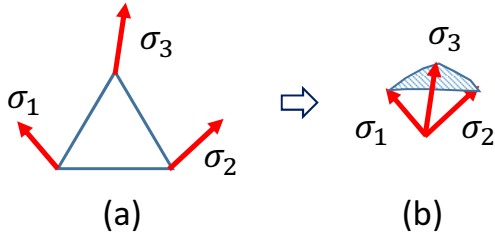}
\caption{(a) Spin variables $\sigma_{i}(i\!=\!1,2,3)$ at the three vertices
of a triangle, and (b) a small triangle area defined by $\sigma_{i}(i\!=\!1,2,3)$
on the unit sphere. This small area can be used to calculate the total
number of skyrmions. }
\label{fig-7} 
\end{figure}

The total number of skyrmions $N_{{\rm sk}}$ is defined by 
\begin{eqnarray}
N_{{\rm sk}}=({1}/{4\pi})\int d^{2}x\;\sigma\cdot\frac{\partial\sigma}{\partial x_{1}}\times\frac{\partial\sigma}{\partial x_{2}},\quad({\rm top})\label{skyrmion-number-top}
\end{eqnarray}
which can be calculated by replacing differentials with differences
\cite{Hog-etal-JMMM2020,Diep-Koibuchi-Frustrated2020}. This $N_{{\rm sk}}$
is denoted by ``top'' and plotted in the figures below. Another
numerical technique for calculating $N_{{\rm sk}}$ is to measure
the solid angle of the triangle cone formed by $\sigma_{1}$, $\sigma_{2}$
and $\sigma_{3}$ (Fig. \ref{fig-7}(a)). Let $a_{\Delta}$ be the
area of the shaded region in Fig. \ref{fig-7}(b), and $N_{{\rm sk}}$
can then be calculated by 
\begin{eqnarray}
N_{{\rm sk}}=\frac{1}{4\pi}\sum_{\Delta}a_{\Delta},\quad({\rm are})\label{skyrmion-number-area}
\end{eqnarray}
and this is denoted by ``are'' below. One more technique to count
$N_{{\rm sk}}$ is denoted by ``gra'', which is a graphical measurement
technique (see Appendix \ref{skx-graphical}).

%f-8
\begin{figure}[h]
\centering{}\includegraphics[width=8.5cm]{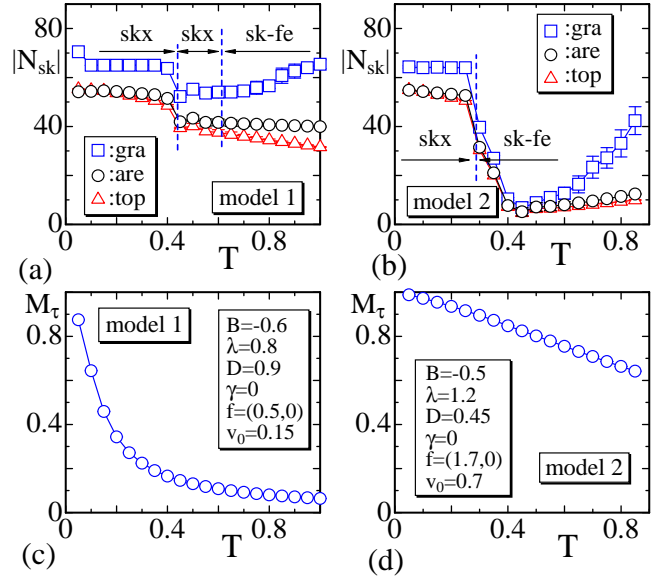}
\caption{Total number of skyrmions $|N_{{\rm sk}}|$ of (a) model 1 and (b)
model 2, where the texts ``gra'', ``are'', and ``top'' correspond
to three different calculation techniques for $|N_{{\rm sk}}|$; ``Gra''
denotes the graphical measurement technique presented in Appendix
\ref{skx-graphical}, ``are'' and ``top'' denote the techniques of
using the formulas in Eqs. (\ref{skyrmion-number-top}) and (\ref{skyrmion-number-area}).
The corresponding order parameter $M_{\tau}$ of (c) model 1 and (d)
model 2 is plotted. }
\label{fig-8} 
\end{figure}

Figure \ref{fig-8}(a) shows the dependence of $|N_{{\rm sk}}|$ of
model 1 on the temperature variation at $B\!=\!-0.6$, where the absolute
values of $N_{{\rm sk}}$ are plotted. These curves in Fig. \ref{fig-8}(a)
are obtained along the horizontal dashed line in Fig. \ref{fig-5}(a).
We find that $|N_{{\rm sk}}|$ discontinuously reduces at $T\!\simeq\!0.45$,
and that the reduced $|N_{{\rm sk}}|$ in the region $T\!>\!0.45$
of ``top'' and ``are''  remain finite up to $T\!\simeq\!1$. Because
of this discontinuous change of $|N_{{\rm sk}}|$, the skx phase of
model 1 is divided into two regions at $T\!\simeq\!0.45$. This skx
phase at higher temperatures is numerically stable. However, $N_{{\rm sk}}$
evaluated graphically, denoted by ``gra'', increases at $T\!\simeq\!0.6$.
This behavior of $N_{{\rm sk}}$ implies that the skx configuration
is collapsed or multiply counted. Therefore, the skx configuration
should be grouped into the sk-fe phase in this region, and we plot
a dashed line as the phase boundary between the skx and sk-fe phases.
The curves $|N_{{\rm sk}}|$ of model 2 in Fig. \ref{fig-8}(b) are
obtained along the horizontal dashed line in Fig. \ref{fig-6}(a)
at $B\!=\!-0.5$, and we find that $N_{{\rm sk}}$ discontinuously
reduces to $N_{{\rm sk}}\!\simeq\!0$. This reduction implies that
the skx phase changes to sk-fe or ferro phase at $T\!\simeq\!0.3$
in model 2.

To see the internal configuration of the 2D non-polar variable $\tau$,
we calculate the order parameter by 
\begin{eqnarray}
M_{\tau}=2\left(\langle\sigma_{x}\rangle^{2}-1/2\right).\label{order-parameter-tau}
\end{eqnarray}
This $M_{\tau}$ continuously changes with respect to $T$ (Fig. \ref{fig-8}(c)
for model 1), and no discontinuous change is observed. However, it
is clear that $\tau$ is anisotropic (isotropic) in the temperature
region $T\!<\!0.2$ ($0.5\!<\!T$). The $M_{\tau}$ plotted in Fig.
\ref{fig-8}(d) for model 2 is very large compared with that in Fig.
\ref{fig-8}(c). This behavior of $M_{\tau}$ implies that $\tau$
is parallel to the direction of $\vec{f}$ in the whole region of
$T$ plotted, resulting from the considerably large value of $f(=\!1.7)$
assumed in model 2 for Fig. \ref{fig-6}.

We should note that the variations of $|N_{{\rm sk}}|$ with respect
to $T$ in Figs. \ref{fig-8}(a) and \ref{fig-8}(b) are identical
to those (which are not plotted) obtained under $\vec{f}\!=\!(0,0)$
and with the same other parameters. In this case, $\gamma$ for $S_{\tau}$
is fixed to $\gamma\!=\!0$, and therefore, $\tau$ becomes isotropic.
This result, obtained under $\vec{f}\!=\!(0,0)$ and $\gamma\!=\!0$,
implies that the skyrmion deformation is caused by the alignment of
$\tau$, and the only effect of $\vec{f}\!\not=\!(0,0)$ is to deform
the skyrmion shape to anisotropic in the skx phase.

%f-9
\begin{figure}[h]
\centering{}\includegraphics[width=8.5cm]{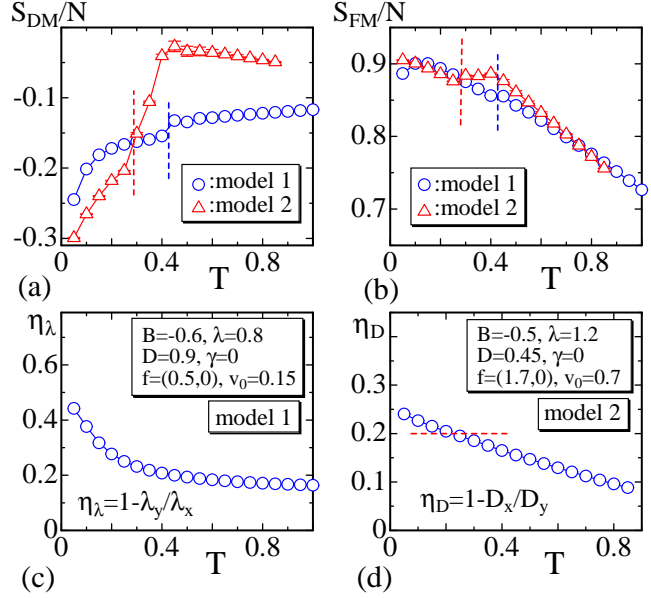}
\caption{(a) $S_{{\rm DM}}/N$ vs. $T$ of model 1 and model 2, (b) $S_{{\rm FM}}/N$
vs. $T$ of model 1 and model 2, the anisotropy of effective interaction
coefficient $\eta_{\lambda}$ and $\eta_{D}$ vs. $T$ of (c) model
1 and (d) model 2. The vertical dashed lines in (a) and (b) roughly
indicate the positions where $S_{{\rm DM}}/N$ and $S_{{\rm FM}}/N$
discontinuously change in model 1 and model 2. The horizontal dashed
line in (d) is drawn at $\eta_{D}\!=\!0.2$, which is the value assumed
in Ref. \cite{Shibata-etal-Natnanotech2015} to simulate the skyrmion
deformation. }
\label{fig-9} 
\end{figure}

The DMI and FMI energies $S_{{\rm DM}}/N$ and $S_{{\rm FM}}/N$ are
shown to have discontinuous changes at $T\!\simeq\!0.4$ in both models
(Figs. \ref{fig-9}(a),(b)), where $N$ is the total number of vertices.
The gaps of these discontinuities in $S_{{\rm FM}}/N$ are very small.

Anisotropy $\eta_{\lambda}$ and $\eta_{D}$ of effective FMI and
DMI coefficients can be evaluated such that 
\begin{eqnarray}
\begin{split} & \eta_{\lambda}=1-\lambda_{y}/\lambda_{x}\quad({\rm model\;1}),\\
 & \eta_{D}=1-D_{x}/D_{y}\quad({\rm model\;2}),
\end{split}
\label{anisotropy-effective-int-coeff}
\end{eqnarray}
where the expressions for $D_{x}$, $D_{y}$ and $\lambda_{x}$, $\lambda_{y}$
are given in Eq. (\ref{anisotropy-effective-D}). The direction dependence
of the definition $\eta_{\lambda}$ of model 1 is different from $\eta_{D}$
of model 2, and this difference comes from the fact that the definition
of $v_{ij}$ in Eq. (\ref{model-1}) for model 1 is different from
that in Eq. (\ref{model-2}) of model 2. We find from the anisotropy
$\eta_{\lambda}$ of model 1 in Fig. \ref{fig-9}(c) that $\eta_{\lambda}$
is decreasing with increasing $T$, and this tendency is the same
for $\eta_{D}$ of model 2 in Fig. \ref{fig-9}(d). It is interesting
to note that $\eta_{D}$ of model 2 is $\eta_{D}\!\simeq\!0.2$ in
the skx phase at $T\!<\!0.4$. This value $\eta_{D}\!=\!0.2$ corresponds
to $D_{x}/D_{y}\!=\!0.8$ explicitly assumed in Ref. \cite{Shibata-etal-Natnanotech2015}
to simulate the skyrmion deformation. This $\eta_{D}$ is slightly
larger than 0.2 at $T\!\simeq0.1$, where the shape anisotropy is
comparable to the experimentally observed one, as demonstrated in
Fig. \ref{fig-6}(d). It must be emphasized that $\eta_{D}$ or equivalently
$D_{x}$ and $D_{y}$ of model 2 are not the input parameters for
the simulations, where the input is $\vec{f}$, and the output is
a skyrmion deformation like in the experiments.

\begin{figure}[h]
\centering{}\includegraphics[width=10.5cm]{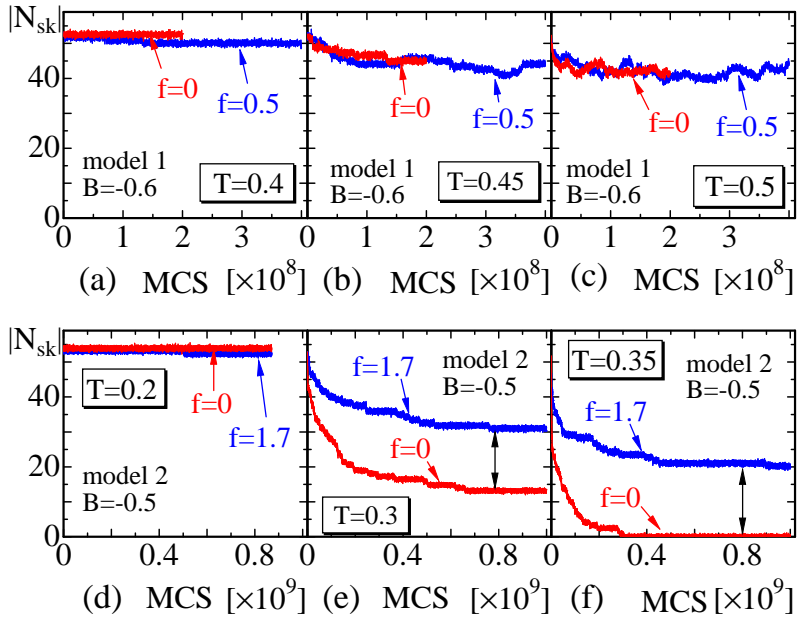}
\caption{
$|N_{\rm sk}|$ vs. MCS obtained on the dashed lines in Figs. \ref{fig-5} and \ref{fig-6} at the boundary between skx and sk-fe phases in (a),(b),(c) model 1 and (d),(e),(f) model 2. $|N_{\rm sk}|$ is independent of whether the stress is applied or not in model 1, while it clearly depends on the stress in model 2. The other parameters $\lambda, D, \gamma, v_0$ are the same  as those shown in Figs. \ref{fig-5} and \ref{fig-6}.
} 
\label{fig-10} 
\end{figure}

Finally in this subsection, we show how the simulations are convergent by plotting $|N_{\rm sk}|$ (top) in Eq. (\ref{skyrmion-number-top}) vs. MCS and discuss how the stress influences the skx phase.  The data $|N_{\rm sk}|$ of model 1 plotted in Figs. \ref{fig-10}(a)--(c), which are obtained on the dashed line in Fig. \ref{fig-5} at the transition region $T\!\simeq\!0.5$, indicate that  the skyrmion number is independent of whether the stress is applied or not. This implies that the distortion of FMI coefficient by uniaxial stress does not influence the skx and sk-fe phases. In contrast, we find in the remaining plots in Figs. \ref{fig-10}(d)--(f), which are obtained on the dashed line in Fig. \ref{fig-6}, that $|N_{\rm sk}|$ of model 2 depends on the stress. Indeed, $|N_{\rm sk}|$ remains unchanged for the stressed condition in the skx phase (Fig. \ref{fig-10}(d)), while $|N_{\rm sk}|$ is considerably increased from $|N_{\rm sk}|\!=\!{\rm finite}$  in the sk-fe phase  (Fig. \ref{fig-10}(e)) and also from $|N_{\rm sk}|\!=\!0$ in the ferro phase  (Fig. \ref{fig-10}(f)).  It is interesting to note that such skyrmion proliferation is experimentally observed by uniaxial stress control not only in low temperature region \cite{Chacon-etal-PRL2015,Nii-etal-PRL2014,Nii-etal-NatCom2015} but also in high temperature region close to the boundary with the ferro phase \cite{Levatic-etal-SCRep2016}. Thus, effects of uniaxial stress on skyrmion proliferation are considered to be implemented in model 2.
%f-10

%----------------------------------------------------------
\subsubsection{Stress vs. magnetic field diagram}
\label{FB-diagram} 
%----------------------------------------------------------
%f-11
\begin{figure}[h]
\centering{}\includegraphics[width=11.5cm]{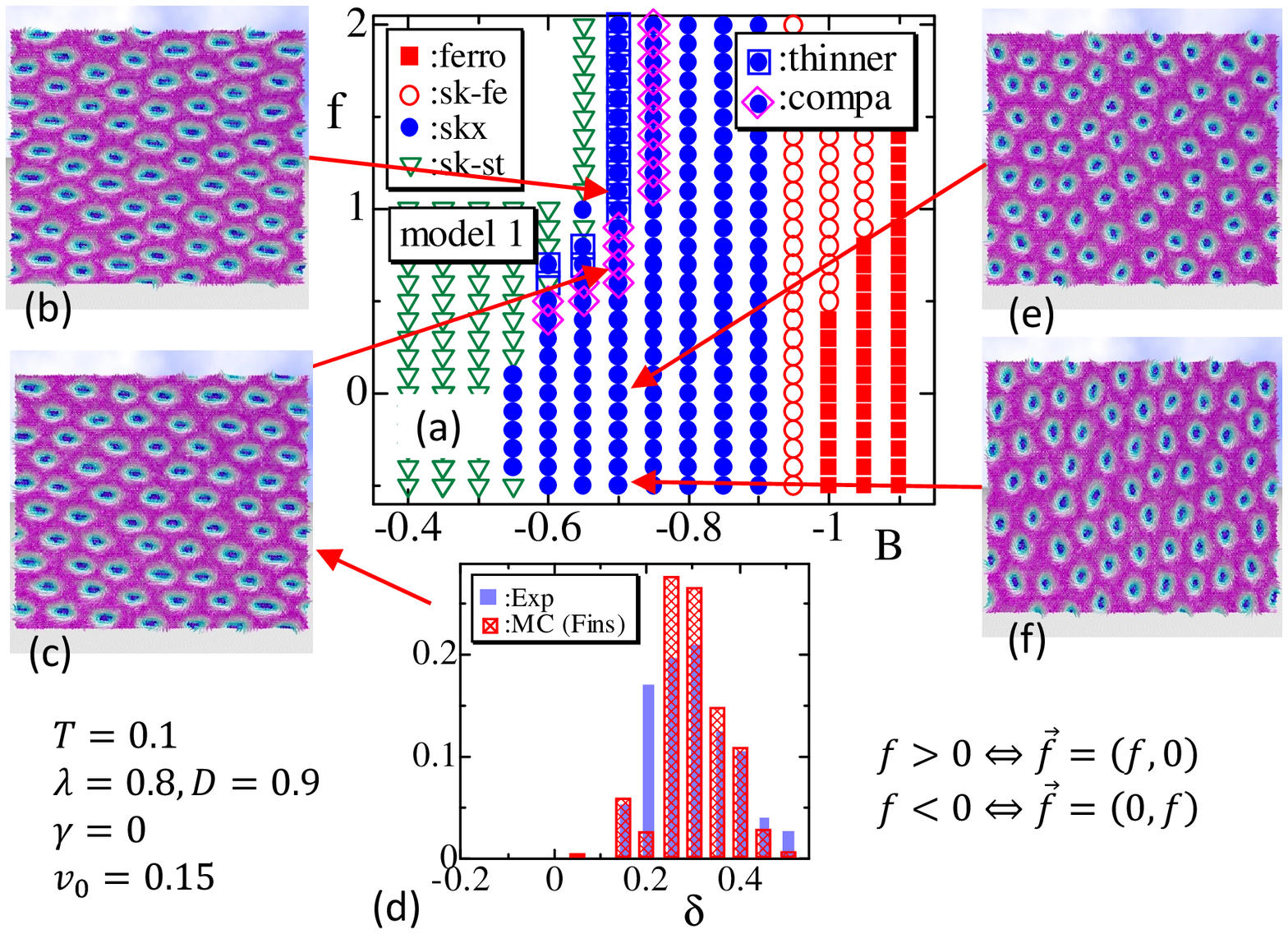}
\caption{(a) $fB$ phase diagram of model 1, where $f$ and $B$ are the external
force and magnetic field, (b)  snapshot of skyrmions at $(f,B)\!=\!(1.1,-0.7)$,
(c)  snapshot obtained at $(f,B)\!=\!(0.7,-0.7)$, (d)  histogram
of $\delta$ corresponding to (c), which is close to Exp data in Ref.
\cite{Shibata-etal-Natnanotech2015}, and (e), (f) snapshots obtained
at $(f,B)\!=\!(0,-0.7)$ and $(f,B)\!=\!(-0.5,-0.7)$, where the negative
$f$ implies ${\vec{f}}\!=\!(0,f)$, and the skyrmion shape deforms
vertically. The assumed parameter values are written in the figure. }
\label{fig-11} 
\end{figure}
%f-12
\begin{figure}[h]
\centering{}\includegraphics[width=11.5cm]{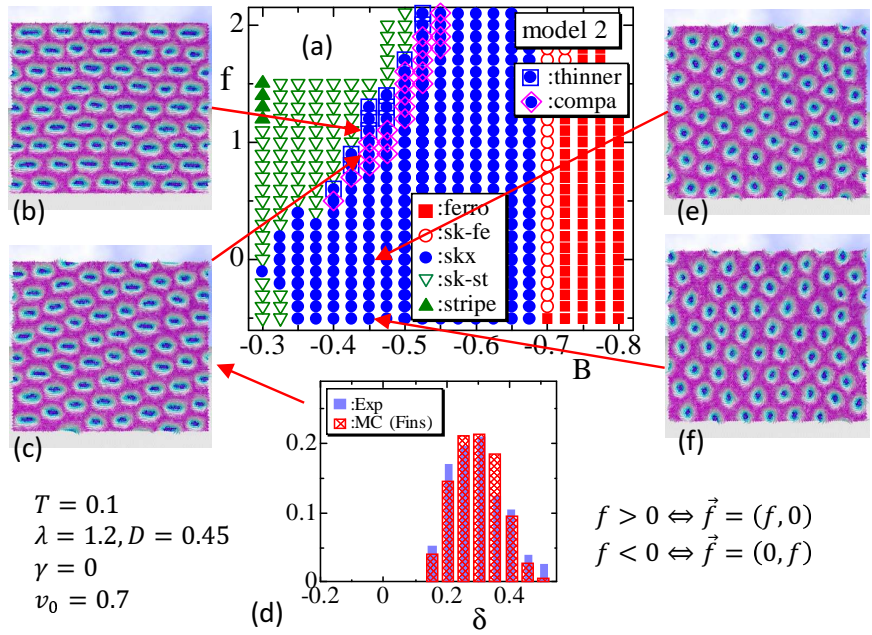}
\caption{(a) $fB$ phase diagram of model 2, where $f$ and $B$ are the external
force and magnetic field, (b)  snapshot of skyrmions at $(f,B)\!=\!(1.1,-0.45)$,
(c)  snapshot obtained at $(f,B)\!=\!(0.9,-0.45)$, (d)  histogram
of $\delta$ corresponding to (c), which is close to Exp data in Ref.
\cite{Shibata-etal-Natnanotech2015}, and (e), (f) snapshots obtained
at $(f,B)\!=\!(0,-0.45)$ and $(f,B)\!=\!(-0.5,-0.45)$, where the
negative $f$ implies ${\vec{f}}\!=\!(0,f)$ and the skyrmion shape
deforms vertically. The assumed parameter values are written on the
figure. }
\label{fig-12} 
\end{figure}

 The external force $f$ and magnetic field $B$ are varied, and $fB$
phase diagrams of model 1 and model 2 are obtained (Figs. \ref{fig-11}
and \ref{fig-12}). The parameters are fixed to $(T,\lambda,D,\gamma,v_{0})\!=\!(0.1,0.8,0.9,0,0.15)$
in Fig. \ref{fig-11} for model 1 and $(T,\lambda,D,\gamma,v_{0})\!=\!(0.1,1.2,0.45,0,0.7)$
in Fig. \ref{fig-12} for model 2. The parameters $(\lambda,D,\gamma,v_{0})$
for model 1 and model 2 are the same as those assumed for the $BT$
phase diagrams in Figs. \ref{fig-5} and \ref{fig-6}. The symbol
(skx) for skyrmion and those for other phases are also exactly the
same as those used in Figs. \ref{fig-5} and \ref{fig-6}.

For the external force $\vec{f}\!=\!(f,0)$ in the positive $x$ direction,
we assign positive $f$ in the vertical axis of the diagrams. In the
case of $\vec{f}\!=\!(f,0)$ for positive $f$, the internal variable
$\tau$ is expected to align along $\vec{f}$ in the direction $(1,0)$
or $x$ direction. In contrast, the negative $f$ in the diagrams
means that $\vec{f}\!=\!(0,f)$. In this case, $\tau$ aligns along
the direction $(0,1)$ or $y$ direction. Such an aligned configuration
of $\tau$ along the $y$ axis is also expected for $\alpha\!=\!-1$
with $\vec{f}\!=\!(f,0)$, because the energy $\alpha S_{f}$ for
$\alpha\!=\!1$ with $\vec{f}\!=\!(0,f)$ is identical to $\alpha S_{f}$
for $\alpha\!=\!-1$ with $\vec{f}\!=\!(f,0)$ up to a constant energy.

Figures \ref{fig-11}(b) and \ref{fig-12}(b) are snapshots of deformed
skyrmions, where the shape anisotropy is slightly larger than the
experimental one in Ref. \cite{Shibata-etal-Natnanotech2015}. In
contrast, the snapshots in Figs. \ref{fig-11}(c) and \ref{fig-12}(c)
are almost comparable in their anisotropy $\delta$, as shown in Figs.
\ref{fig-11}(d) and \ref{fig-12}(d) with the experimentally reported
one denoted by Exp. The word ``thinner'' corresponding to the solid
circle enclosed by a blue-colored square indicates that the shape
deformation is thinner than that of Exp, and the word ``compa''
corresponding to that enclosed by a pink-colored diagonal indicates
that the shape deformation is comparable to that of Exp. 
For $f\!=\!0$, the skyrmion shape is isotropic, as we see in Figs.
\ref{fig-11}(e) and \ref{fig-12}(e), and the shape vertically deforms
for the negative $f$ region, which implies positive $f$ in $\vec{f}\!=\!(0,f)$,
in Figs. \ref{fig-11}(f) and \ref{fig-12}(f). Thus, we can confirm
from the snapshots that the shape deforms to oblong along the applied
tensile force direction. Moreover, the deformation is almost the same
as Exp for a certain range of $f$ in both model 1 and model 2.

We should note that the skx phase changes to the sk-st phase with
increasing $f$ at a relatively small $B$ region, however, it does
not change to the ferro phase at an intermediate region of $B$ even
if $f$ increases to sufficiently large, where $\tau$ saturates in
the sense that no further change is expected. This saturation is because
the role of $\vec{f}$ is only to rotate the direction of $\tau$.
Hence, the modeling of stress by $\vec{f}$ and $\tau$ is considered
effective only in small stress regions, as mentioned in Section \ref{models}.
This point is different from the reported numerical results in Ref.
\cite{JWang-etal-PRB2018}, where the skx phase terminates, and the
stripe or ferro phase appears for sufficiently large strain in the
strain vs. magnetic field diagram.

\begin{figure}[h]
\centering{}\includegraphics[width=11.5cm]{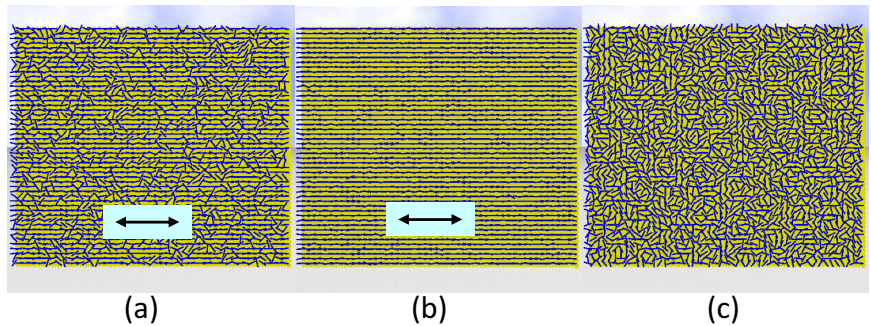}
\caption{Snapshots of $\tau$ corresponding to (a) Fig. \ref{fig-5}(c) of
model 1, (b) Fig. \ref{fig-6}(c) of model 2, and (c) Fig. \ref{fig-12}(e)
of model 2. The small cylinders correspond to $\tau$. The total number
of  cylinders is reduced to 2500, which is quarter  of $N(=\!10000)$,
to clarify the directions. The arrows ($\leftrightarrow$) in (a)
and (b) denote the direction of tensile force $\vec{f}$. }
\label{fig-13} 
\end{figure}

Finally in this subsection, we show snapshots of the variable $\tau$
in Figs. \ref{fig-13}(a), (b), and (c), which  correspond to the
configurations shown in Fig. \ref{fig-5}(c) of model 1, Fig. \ref{fig-6}(c)
of model 2, and Fig. \ref{fig-12}(e) of model 2, respectively. To
clarify the directions of $\tau$, we show a quarter of $\tau$ ($\Leftrightarrow$ the total number of $\tau$ is 2500) in the snapshots. We find that $\tau$ is almost parallel to $\vec{f}$ denoted by the arrows in
(a) and (b), and it is almost random in (c), where $f$ is assumed
to be $f\!=\!0$. The reason why $\tau$ in (b) is more uniform than
in (a) is because $f(=\!1.7)$ in (b) is relatively larger than $f(=\!0.5)$
in (a). 

% f-13

%----------------------------------------------------------
\subsection{Responses to uniaxial strains}
\label{uniaxial-strains} 
%----------------------------------------------------------
To summarize the results in Section \ref{uniaxial-stress}, both model
1 and model 2 successfully describe the shape deformation of skyrmions
under external mechanical forces $\vec{f}$. The skyrmion deformation
comes from the fact that the skx phase is sensitive to the direction
$\tau$ of the strain field influenced by $f$ in $\vec{f}\!=\!(f,0)$,
which is assumed to be positive or equivalently tensile, as mentioned
in Section \ref{models}. This successful result implies that the
interaction between spins and the mechanical force is adequately implemented
in both models at least in the skx phase.

Besides, the response of spins in the stripe phase in both models,
or more explicitly, the stripe direction as a response to $\vec{f}$
is also consistent with the reported experimental result in Ref. \cite{JDho-etal-APL2003}.
In this Ref. \cite{JDho-etal-APL2003}, as mentioned in the introduction,
Dho et al. experimentally studied magnetic microstructures of LSMO
thin film at room temperature and zero magnetic fields. The reported
results indicate that the direction of the strain-induced magnetic
stripe becomes dependent on whether the force is compression or tension.

On the other hand, the definition of $S_{{\rm DM}}$ in Eqs. (\ref{model-1}),
(\ref{model-2}) is explicitly dependent on the shape of the lattice,
and therefore, we examine another check for the response of spins
in the stripe phase by deforming the lattice itself, as described
in Fig. \ref{fig-3}. To remove the effect of $\vec{f}$, we fix $\vec{f}$
to $f\!=\!0$ in $S_{f}$, and instead, $\gamma$ in $\gamma S_{\tau}$
is changed from $\gamma\!=\!0$ to $\gamma\!=\!0.5$ for model 1 and
$\gamma\!=\!0.65$ for model 2. As a consequence of these non-zero
$\gamma$, the variable $\tau$ is expected to align to some spontaneous
directions. If the lattice deformation non-trivially influences $\tau$,
this spontaneously and locally oriented configuration of $\tau$ is
expected to influence spin configurations strongly in the stripe phase.
As a consequence, the stripe direction becomes anisotropic on deformed
lattices ($\Leftrightarrow\xi\!\not=\!1$), while the stripe is isotropic
on the undeformed lattice ($\Leftrightarrow\xi\!=\!1$).

To check these expectations by the lattice deformations shown in Fig.
\ref{fig-3}, we modify the unit tangential vector ${\vec{e}}_{ij}$,
which originally comes from $\partial{\vec{r}}_{i}/\partial x_{j}$
(Appendix \ref{FG-model}). Indeed, $\partial{\vec{r}}_{i}/\partial x_{j}$
is understood to be the edge vector ${\vec{\ell}}_{ij}(=\!\vec{r}_j\!-\!\vec{r}_i)$ from vertex
$i$ to vertex $j$ in the discrete model, and therefore, both the
direction and the length of ${\vec{\ell}}_{ij}$ are changed by the
lattice deformations in Fig. \ref{fig-3}. Thus, the unit tangential
vector ${\vec{e}}_{ij}\!=\!(e_{ij}^{x},e_{ij}^{y})$ in $S_{{\rm DM}}$
in Eqs. (\ref{model-1}) and (\ref{model-2}) is replaced by 
\begin{eqnarray}
{\vec{e}}_{ij}^{\;\prime}=(e_{ij}^{\prime x},e_{ij}^{\prime y})=(\xi^{-1}e_{ij}^{x},\xi e_{ij}^{y}).\label{new-bond-vect}
\end{eqnarray}
This generalized vector ${\vec{e}}_{ij}^{\;\prime}$ is identical
to the original unit vector ${\vec{e}}_{ij}$ for $\xi\!=\!1$.
 Note also that ${\vec{e}}_{ij}$ in $v_{ij}$ in Eqs. (\ref{model-1})
and (\ref{model-2}) is  replaced by ${\vec{e}}_{ij}^{\;\prime}$ as follows: 
\begin{eqnarray}
\begin{split} & S_{{\rm FM}}=\sum_{\Delta}\left[\lambda_{ij}\left(1-\sigma_{i}\cdot\sigma_{j}\right)+\lambda_{jk}\left(1-\sigma_{j}\cdot\sigma_{k}\right)+\lambda_{ki}\left(1-\sigma_{k}\cdot\sigma_{i}\right)\right],\\
 & S_{{\rm DM}}=\sum_{ij}{\vec{e}}_{ij}^{\;\prime}\cdot\sigma_{i}\times\sigma_{j},  \\
 & \lambda_{ij}=\frac{1}{3}\left(\frac{v_{ij}}{v_{ik}}+\frac{v_{ji}}{v_{jk}}\right),\quad v_{ij}=\left\{ \begin{array}{@{\,}ll}
                 |\tau_{i}\cdot{\vec{e}}_{ij}^{\;\prime}|+v_{0} &  (|\tau_{i}\cdot{\vec{e}}_{ij}^{\;\prime}|<1) \\
                 1+v_0 &  (|\tau_{i}\cdot{\vec{e}}_{ij}^{\;\prime}|\geq 1) 
                  \end{array} 
                   \right., \quad({\rm model\;1}), 
\end{split}
\label{model-1prime}
\end{eqnarray}
and 
\begin{eqnarray}
\begin{split} & S_{{\rm FM}}=\sum_{ij}\left(1-\sigma_{i}\cdot\sigma_{j}\right),\\
 & S_{{\rm DM}}=\sum_{\Delta}\left[\lambda_{ij}\left({\vec{e}}_{ij}^{\;\prime}\cdot\sigma_{i}\times\sigma_{j}\right)+\lambda_{jk}\left({\vec{e}}_{jk}^{\;\prime}\cdot\sigma_{j}\times\sigma_{k}\right)+\lambda_{ki}\left({\vec{e}}_{ki}^{\;\prime}\cdot\sigma_{k}\times\sigma_{i}\right)\right],\\
 & \lambda_{ij}=\frac{1}{3}\left(\frac{v_{ij}}{v_{ik}}+\frac{v_{ji}}{v_{jk}}\right),\quad v_{ij}=\left\{ \begin{array}{@{\,}ll}
                 \sqrt{1-\left(\tau_{i}\cdot{\vec{e}}_{ij}^{\;\prime}\right)^{2}}+v_{0} &  (|\tau_{i}\cdot{\vec{e}}_{ij}^{\;\prime}|<1) \\
                 v_0 &  (|\tau_{i}\cdot{\vec{e}}_{ij}^{\;\prime}|\geq 1) 
                  \end{array} 
                   \right., \quad({\rm model\;2}),  
\end{split}
\label{model-2prime}
\end{eqnarray}
and the corresponding models are also denoted by model 1 and model 2. The difference between models in Eqs. (\ref{model-1prime}), (\ref{model-2prime}) and Eqs. (\ref{model-1}), (\ref{model-2}) comes from the definition of $v_{ij}$. However, the variables $v_{ij}$ in  Eqs. (\ref{model-1prime}), (\ref{model-2prime})  are identical with $v_{ij}$ in Eqs. (\ref{model-1}), (\ref{model-2}) for the non-deformed lattice corresponding to $\xi\!=\!1$, and therefore, both models in Eqs. (\ref{model-1prime}), (\ref{model-2prime}) are simple and straightforward extension of models in Eqs. (\ref{model-1}), (\ref{model-2}). 
 From the definitions of  $v_{ij}$ in Eqs. (\ref{model-1prime}) and (\ref{model-2prime}), $v_{ij}$ no longer have the meaning of a component of $\tau_i$ along or perpendicular to the direction from vertex $i$ to vertex $j$.  It is also possible to start with model 1 and model 2 in Eqs. (\ref{model-1prime}) and (\ref{model-2prime}) from the beginning, however, model 1 and model 2 in Eqs. (\ref{model-1}), (\ref{model-2}) are relatively simple and used to study responses to the external stress $\vec f$ in Section \ref{uniaxial-stress}.

\begin{figure}[h]
\centering{}\includegraphics[width=11.5cm]{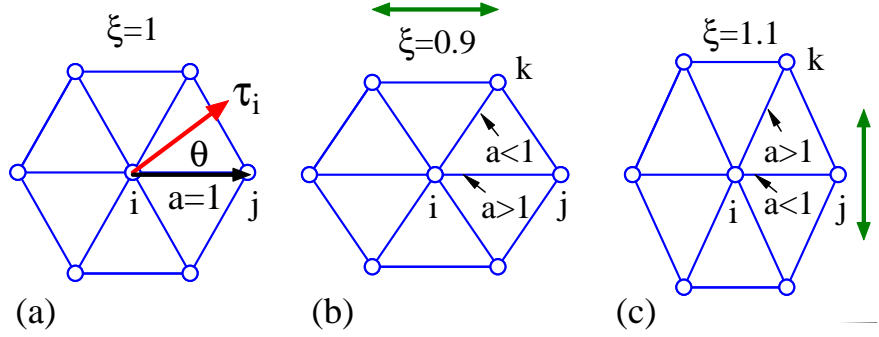}
\caption{
(a) Regular triangular lattice corresponding to $\xi\!=\!1$, and deformed lattices corresponding to (b) $\xi\!=\!0.9$ and (c) $\xi\!=\!1.1$. The bond length $a$ in (a) is $a\!=\!1$, while in (b) and (c), $a$ changes to $a>1$ or $a<1$ depending on the direction of bonds. The symbol $\theta$ in (a) is the angle between $\tau_i$ and the direction of bond $ij$, and the arrows ($\leftrightarrow$) and ($\updownarrow$) in (b) and (c) indicate the elongation direction.
} 
\label{fig-14} 
\end{figure}

 Since the definition of $v_{ij}$ in Eqs. (\ref{model-1prime}) and (\ref{model-2prime}) depends on the bond vector ${\vec{e}}_{ij}^{\;\prime}$, we first show the lattices corresponding to $\xi\!=1$,  $\xi\!=0.9$, and $\xi\!=1.1$ in Figs.  \ref{fig-14}(a)--(c). 
Let the bond length or the lattice spacing $a(=\!|{\vec{e}}_{ij}|)$ be $a\!=\!1$ on the regular lattice, then $a(=\!|{\vec{e}}_{ij}^{\;\prime}|)$ becomes $a>1$ or $a<1$ depending on the bond direction on the deformed lattices. For $\xi\!=\!0.9$, all bonds in the horizontal direction,  such as bond $ij$ in Fig. \ref{fig-14}(b), satisfy $a>1$, and all other bonds, such as bond $ik$, satisfy $a<1$. To the contrary, for $\xi\!=\!1.1$, all bonds in the horizontal direction satisfy $a<1$ and all other bonds satisfy $a>1$ as shown in Fig. \ref{fig-14}(c).

%f-14

%f-15
\begin{figure}[h]
\centering{}\includegraphics[width=11.5cm]{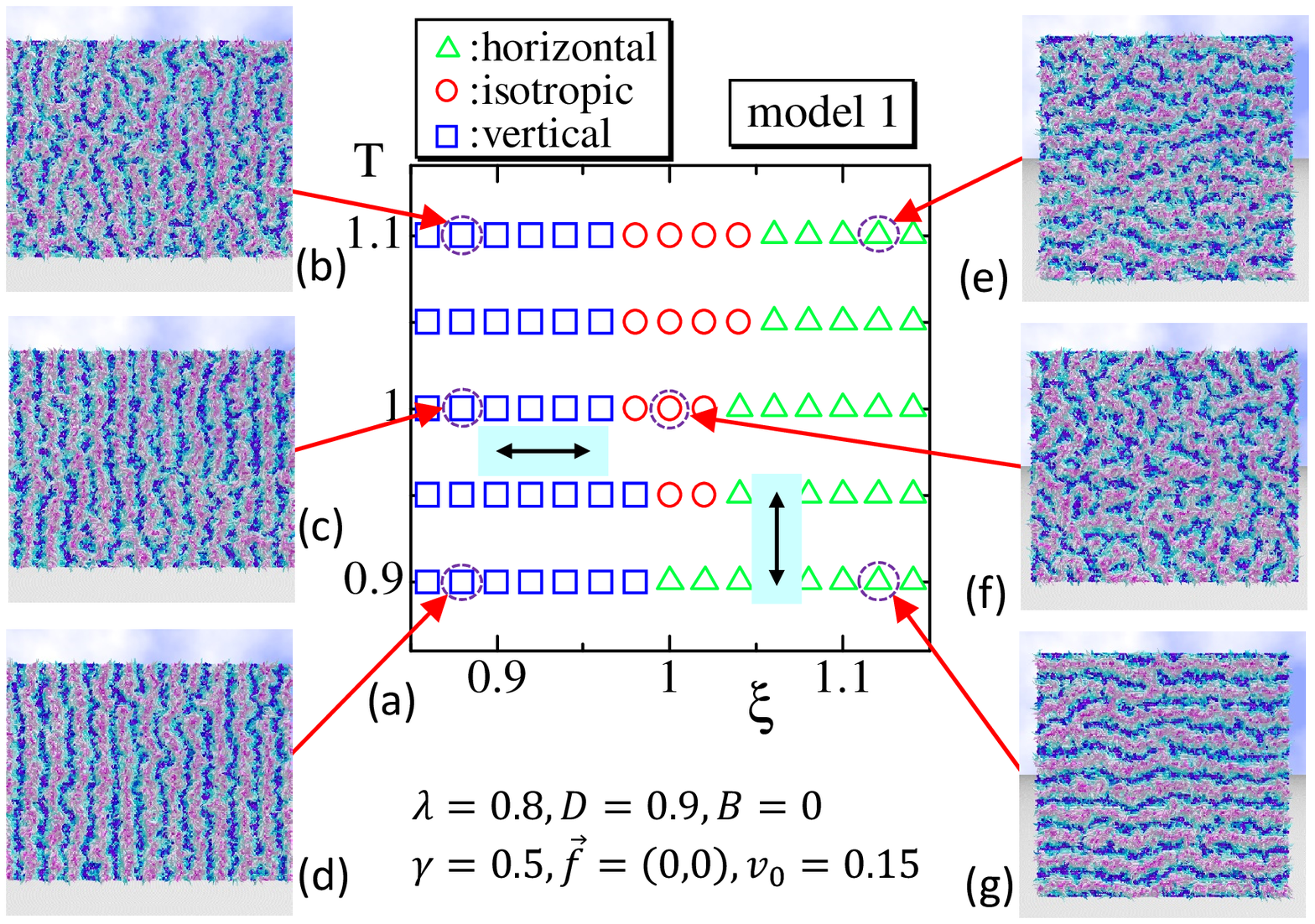}
\caption{(a) $T\xi$ diagram in the stripe phase of model 1, where $T$ and
$\xi$ are the temperature and deformation parameter in Eq. (\ref{strain-deformation}).
The arrows ($\leftrightarrow$) and ($\updownarrow$) denote the lattice
elongation direction, whereas the symbols (\textcolor{green}{$\bigtriangleup$}),
(\textcolor{red}{$\bigcirc$}) and (\textcolor{blue}{$\square$})
denote alignments of the stripe direction. (b), (c) and (d) are snapshots
obtained at $\xi\!=\!0.88$, and (e), (f) and (g) are those obtained
at $\xi\!=\!1$ and $\xi\!=\!1.12$. The parameters $\lambda$ and
$D$ are the same as those used in Figs. \ref{fig-5} and \ref{fig-11},
and $(B,\gamma,f)$ are fixed to $(B,\gamma,f)\!=\!(0,0.5,0)$. Fluctuations
of spins increase with increasing temperature. }
\label{fig-15} 
\end{figure}

%f-16
\begin{figure}[h]
\centering{}\includegraphics[width=11.5cm]{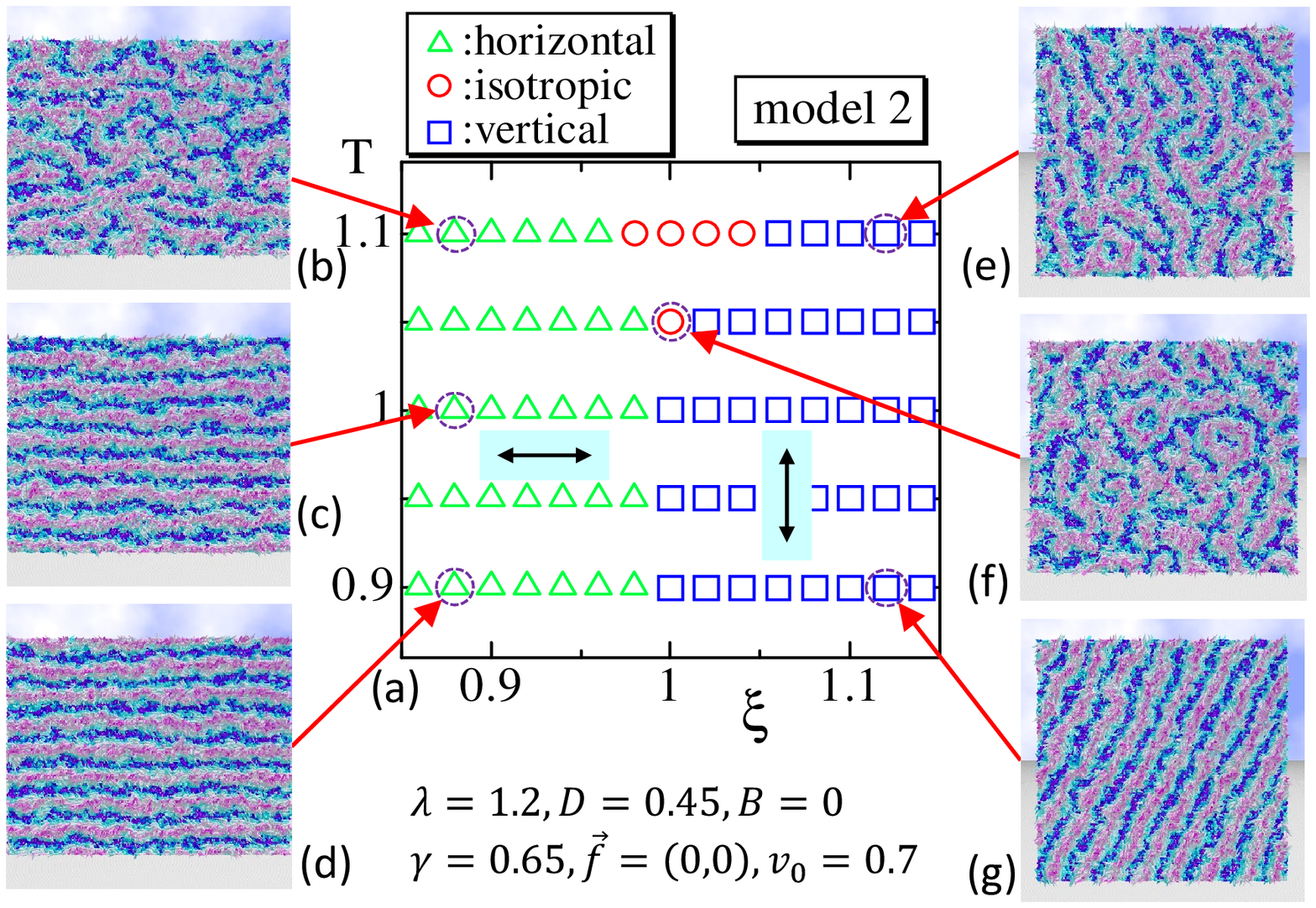}
\caption{(a) $T\xi$ diagram in the stripe phase of model 2, where $T$ and
$\xi$ are the temperature and deformation parameter in Eq. (\ref{strain-deformation}).
The arrows ($\leftrightarrow$) and ($\updownarrow$) denote the lattice
elongation direction, whereas the symbols (\textcolor{green}{$\bigtriangleup$}),
(\textcolor{red}{$\bigcirc$}),and (\textcolor{blue}{$\square$})
denote alignments of the stripe direction. (b), (c) and (d) are snapshots
obtained at $\xi\!=\!0.88$, and (e), (f) and (g) are those obtained
at $\xi\!=\!1$ and $\xi\!=\!1.12$. The parameters $\lambda$ and
$D$ are the same as those used in Figs. \ref{fig-6} and \ref{fig-12},
and $(B,\gamma,f)$ are fixed to $(B,\gamma,f)\!=\!(0,0.65,0)$. Fluctuations
of spins increase with increasing temperature. }
\label{fig-16} 
\end{figure}

We should comment on the influences of lattice deformation described
in Eq. (\ref{strain-deformation}) on $S_{{\rm FM}}$ and $S_{{\rm DM}}$
in model 1 and model 2 in detail. First, the definition of $S_{{\rm DM}}$
initially depends on the lattice shape. Moreover, in $S_{{\rm DM}}$
of model 2, the influences of lattice deformation come from both ${\vec{e}}_{ij}^{\;\prime}$
and $\lambda_{ij}$, which depends on $v_{ij}$. $S_{{\rm FM}}$ in
model 1 is also dependent on the lattice shape due to this $\lambda_{ij}$.
In contrast, $S_{{\rm FM}}$ in model 2 depends only on the connectivity
of the lattice and is independent of the lattice shape. To summarize,
the lattice deformation by $\xi$ in Eq. (\ref{strain-deformation})
influences both $S_{{\rm FM}}$ and $S_{{\rm DM}}$ in model 1, and
it influences only $S_{{\rm DM}}$ in model 2.

Figures \ref{fig-15} and \ref{fig-16} show phase diagrams for the
stripe phase in model 1 and model 2 under  variations of $\xi$ and
$T$. The symbols (\textcolor{green}{$\bigtriangleup$}), (\textcolor{red}{$\bigcirc$}),and
(\textcolor{blue}{$\square$}) denote horizontal, isotropic, and vertical
alignments of stripe direction. In Fig. \ref{fig-16} (g), the alignment
direction is not exactly vertical to the horizontal direction, but
it is parallel to the triangle's edge directions (see Fig. \ref{fig-1}).
This deviation in the alignment direction is in contrast to the case
of model 1 in Figs. \ref{fig-15}(b), (c) and(d) and is also in contrast
to the case that $\vec{f}\!=\!(0,f)$ is applied, where the stripe
direction is precisely vertical to the horizontal direction (which
is not shown). For $\xi\!=\!1$, the lattice is not deformed, and
uniaxial strains, and hence, aligned stripes are not expected. Indeed,
we find from the snapshots in Figs. \ref{fig-15} and \ref{fig-16}
that the stripe direction is not always uniformly aligned, except
for at relatively low temperatures such as $T\!=\!0.9$. From this,
it is reasonable to consider $\tau$ to be a strain direction in a
microscopic sense. If $\gamma$ is fixed to a larger value, such as
$\gamma\!=\!1$ in both models, then the stripe pattern, or equivalently
the direction of $\tau$, becomes anisotropic even at $\xi\!=\!1$
like those in the case of $\xi\!\not=\!1$.

We find that the results of model 2 in Fig. \ref{fig-16} are consistent
with the reported experimental data in Ref. \cite{JDho-etal-APL2003}
(see Fig. \ref{fig-3}), implying that $\tau$ in model 2 correctly
represents the direction of strains expected under the lattice deformations
by $\xi$. On the contrary, the results of model 1 in Fig. \ref{fig-15}
are inconsistent with the experimental data. This difference in the
stripe direction comes from the fact that the lattice deformation
incorrectly influences the alignment of $\tau$, or in other words,
$\tau$ in model 1 is not considered as the strain direction corresponding
to the lattice deformation.

Thus, the strains caused by lattice deformations are consistent (inconsistent)
to their stress type, compression, or tension, which determines the
direction of stripe pattern in model 2 (model 1) at $T\!\simeq\!1$
and $B\!=\!0$. For the low-temperature region, the responses of lattice
deformation in model 2 and model 1 are partly inconsistent with the
experimental result in Ref. \cite{JDho-etal-APL2003}. To summarize,
the numerical results in this paper support that the reason for skyrmion
shape deformation, described in Ref. \cite{Shibata-etal-Natnanotech2015},
is an anisotropy in the DMI coefficient.

% f-17
\begin{figure}[h]
\centering{}\includegraphics[width=11.5cm]{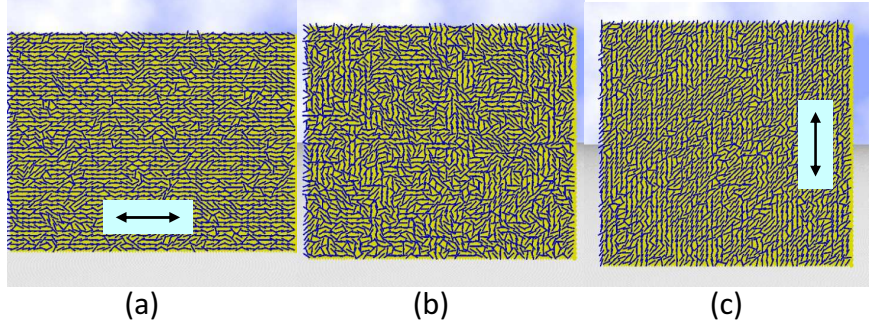}
\caption{Snapshots of $\tau$ of model 2 obtained at (a) $(T,B)\!=\!(0.9,0.88)$,
(b) $(T,B)\!=\!(1.05,1)$, and (c) $(T,B)\!=\!(0.9,1.12)$, which
 correspond to Figs. \ref{fig-16}(d), \ref{fig-16}(f), and \ref{fig-16}(g),
respectively. The small cylinders represent $\tau$. The total number
of  cylinders is reduced to 2500, which is quarter of $N(=\!10000)$,
to clarify the directions. The arrows in (a) ($\leftrightarrow$)
and (c) ($\updownarrow$) denote the lattice elongation directions. }
\label{fig-17} 
\end{figure}

Here we show snapshots of $\tau$ in Figs. \ref{fig-17}(a), (b) and
(c) corresponding to Figs. \ref{fig-16}(d), \ref{fig-16}(f), and
\ref{fig-16}(g), respectively. We find that almost all $\tau$ align
along the horizontal direction in (a), the direction locally aligns
and is globally isotropic in (b), and almost all $\tau$ align along
the vertical direction or the triangle edge direction in (c).
The random state of $\tau$ in Fig. \ref{fig-17}(b) implies that the direction of $D$-vector is globally 
at random and considered to correspond to a non-coplanar distribution of $D$-vectors in the bulk system with inhomogeneous distortion expected from the effective magnetic model \cite{Plumer-Walker-JPC1982,Plumer-etal-JPC1984}. Thermal fluctuations in such a random state may grow on larger lattices, and if such an unstable phenomenon is expected, the deformed skyrmion shape changes with increasing lattice size. However,  no difference is found in the simulation results on the lattice of size $100\!\times\! 100$ and those on the lattices of $200\!\times\! 200$ and $400\!\times\! 400$ on the dashed lines in Figs. \ref{fig-5} and \ref{fig-6}. Due to the competing interactions in our model, the spin configuration is non-uniform with topological textures. However, skyrmion structures cannot be generated by random anisotropies.

% f-18
\begin{figure}[h]
\centering{}\includegraphics[width=11.5cm]{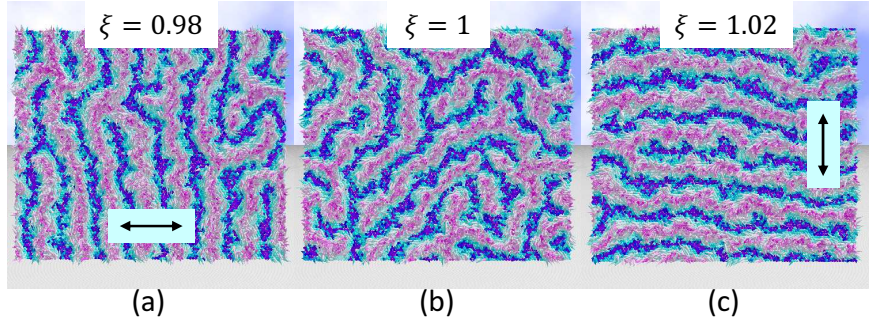}
\caption{Responses of the original model, in which the FG prescription is not applied, to the lattice deformations (a) $\xi\!=\!0.98$,
(b) $\xi\!=\!1$ and (c) $\xi\!=\!1.02$ in the stripe phase for $(T,\lambda,D,B)\!=\!(1,1.6,0.9,0)$.
The direction of stripes for $\xi\!\protect\not=\!1$ is inconsistent
with the experimental result in Ref. \cite{JDho-etal-APL2003}. The
arrows inside the snapshots of (a) ($\leftrightarrow$) and (c) ($\updownarrow$)
denote the lattice elongation direction. }
\label{fig-18} 
\end{figure}

To further check the response of spins to the lattice deformation,
we examine the original model defined by \cite{Hog-etal-JMMM2020,Diep-Koibuchi-Frustrated2020}
\begin{eqnarray}
\begin{split} & S=\lambda S_{{\rm FM}}+DS_{{\rm DM}}-S_{B},\\
 & S_{{\rm FM}}=\sum_{ij}\left(1-\sigma_{i}\cdot\sigma_{j}\right),\quad S_{{\rm DM}}=\sum_{ij}{\vec{e}}_{ij}^{\;\prime}\cdot\sigma_{i}\times\sigma_{j},
\end{split}
\label{original-Hamiltonian}
\end{eqnarray}
where both $S_{{\rm FM}}$ and $S_{{\rm DM}}$ are not deformed by
FG modeling prescription, and $S_{B}$ is the same as in Eq. (\ref{other-Hamiltonian}).
The $S_{{\rm DM}}$ is defined by using the generalized ${\vec{e}}_{ij}^{\;\prime}$
in Eq. (\ref{new-bond-vect}). The parameters are assumed as $(T,\lambda,D,B)\!=\!(1,1.6,0.9,0)$.
The snapshots are shown in Figs. \ref{fig-18}(a), (b) and (c) for
$\xi\!=\!0.98$, $\xi\!=\!1$, and $\xi\!=\!1.02$, respectively.
We find that the result is inconsistent with the reported experimental
data in Ref. \cite{JDho-etal-APL2003}. This inconsistency implies
that the effective coupling constants, such as $D_{x}$ and $D_{y}$
in Eq. (\ref{anisotropy-effective-D}), play a non-trivial role in
the skyrmion deformation and the stripe direction. It must also be
emphasized that stress-effect implemented in model 2 via the alignment of $\tau$
correctly influences helical spin configurations of the skyrmion shape deformation
and the stripe direction.

For smaller (larger) $\xi$, such as $\xi\!=\!0.94$ ($\xi\!=\!1.06$),
the vertical (horizontal) direction of stripes becomes more apparent
in Fig. \ref{fig-18}. Interestingly, the vertical direction of stripes
is parallel to the $y$ direction and not parallel to the triangle
edge directions. This result indicates that the vertical direction
shown in Fig. \ref{fig-16}(g) comes from non-trivial effects of $\lambda_{ij}$
of $S_{{\rm FM}}$ and $S_{{\rm DM}}$ in Eqs.  (\ref{model-1prime}) and (\ref{model-2prime}). We note that the parameters are not always limited
to those used in Fig. \ref{fig-18}. It is possible to use a wide
range of $(T,\lambda,D)$ where isotropic stripe configurations like
in Fig. \ref{fig-18}(b) are expected for $\xi\!=\!1$.

%f-19
\begin{figure}[h]
\centering{}\includegraphics[width=8.5cm]{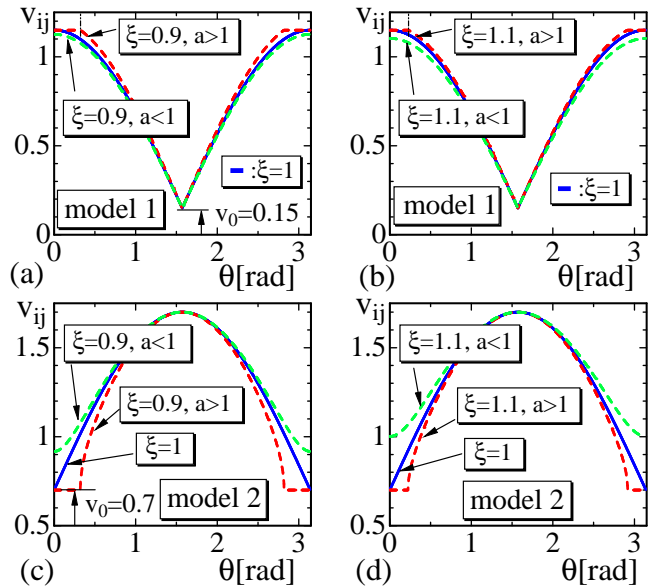}
\caption{
The variation of $v_{ij}$ vs. $\theta$ of model 1 for (a) $\xi\!=\!0.9$ and (b)  $\xi\!=\!1.1$,  and $v_{ij}$ vs. $\theta$ of model 2 for (c) $\xi\!=\!0.9$ and (d)  $\xi\!=\!1.1$, where $\theta$ is the angle between $\tau_i$ and ${\vec{e}}_{ij}^{\;\prime}$ (see Fig. \ref{fig-14}(a)). All the curves of $v_{ij}$ (dashed lines) continuously reduce to the curve of $v_{ij}$ (solid line) in the limit of $\xi\!\to\!1$. 
} 
\label{fig-19} 
\end{figure}
 In Fig. \ref{fig-19}(a), the Finsler length $v_{ij}$ defined by Eq. (\ref{model-1prime}) for $\xi\!=\!0.9$ are plotted, where the horizontal axis $\theta$ is the angle between $\tau_i$ and ${\vec{e}}_{ij}^{\;\prime}$ (see Fig. \ref{fig-14}(a)). For $\xi\!=\!1$,  $v_{ij}$ (dashed line) is identical with the original $v_{ij}$ in Eq. (\ref{model-1}), which is also plotted (solid line) and is found to be shifted from $[0,1]$ to $[v_0,1\!+\!v_0]$ by a constant $v_0(=\!0.15)$ in Figs. \ref{fig-19} (a),(b) and $v_0(=\!0.7)$ in Figs. \ref{fig-19} (c),(d). We find that $v_{ij}$ (dashed line) deviates from $v_{ij}$ (solid line) only slightly at the region $\theta\to 0$ or equivalently $\theta\to \pi$, while at $\theta\!\to\! \pi/2$, $v_{ij}$ (dashed line) is identified with $v_{ij}$ (solid line) for any $\xi$. On the lattice of $\xi\!=\!1.1$ in Fig. \ref{fig-19}(b), the behavior of $v_{ij}$ (dashed line) is almost comparable to the case of $\xi\!=\!0.9$ in Fig. \ref{fig-19}(a). In addition, the curve of $v_{ij}$ (dashed line) of model 2 on the long bond $a\!>\!1$ also includes a constant part ($=\!v_0$) at $\theta\!\to\!0$ like in the case of model 1. This constant part disappears in the limit of $\xi\!\to\!1$, and hence,  model 2 in Eq. (\ref{model-2prime})   as well as model 1 in Eq. (\ref{model-1prime}) is understood to be an extension of those in Eq. (\ref{model-1}) and (\ref{model-2}) as mentioned above.

%f-20
\begin{figure}[h]
\centering{}\includegraphics[width=10.5cm]{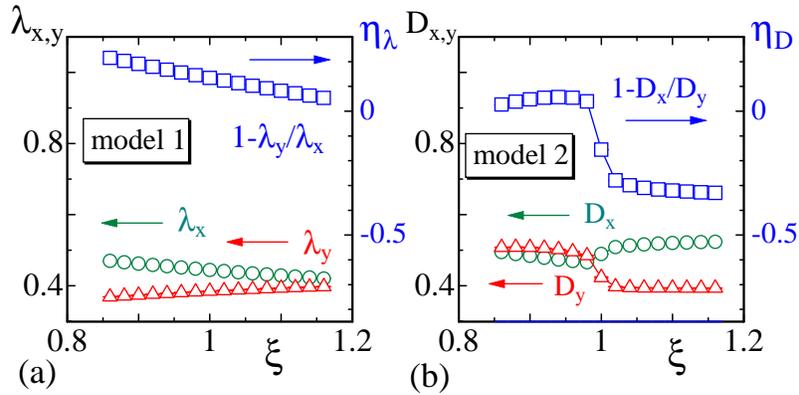}
\caption{
(a) The effective coupling constants $\lambda_x, \lambda_y$ and the anisotropy $\eta_\lambda$ vs. $\xi$ of model 1, and (b) $D_x, D_y$ and $\eta_D$ of model 2. The behaviors of $\lambda_\mu$ and $\eta_\lambda$ of model 1 are almost identical to those of model 2 except the jumps at $\xi\!=\!1$ in model 2. These data of model 1 (model 2) are obtained from the simulations in Fig. \ref{fig-15} (Fig. \ref{fig-16}) at $T\!=\!1$. 
}
\label{fig-20} 
\end{figure}
Now, we discuss why the results of model 2 are considered to be  more realistic. We show the variation of effective coupling constants $\lambda_\mu$ and $D_\mu$ ($\mu\!=\!x,y$), defined by Eq. (\ref{anisotropy-effective-D}), with respect to $\xi$ (Figs. \ref{fig-20}(a),(b)), where the anisotropies $\eta_\lambda$ and $\eta_D$, defined by Eq. (\ref{anisotropy-effective-int-coeff}), are also plotted. We find that in the region $\xi\!>\!1$, both $\eta_\lambda$ and $\eta_D$ are decreasing and smaller than those in $\xi\!<\!1$ in Figs. \ref{fig-20}(a),(b).  Remarkably, the variations of $D_x$, $D_y$ and $\eta_D$ vs. $\xi$ in model 2 almost discontinuously change at $\xi\!=\!1$ and are in sharp contrast to those of model 1. In model 2, if $\eta_D$ is positive (negative), which implies $D_x\!<\!D_y$ ($D_x\!>\!D_y$), then the stripe direction is horizontal (vertical).  Thus, we find that model 2 on the deformed lattices for $\xi\!<\!1$ (Fig. \ref{fig-14}(b)) shares the same property as that on the non-deformed lattice with a tensile stress ${\vec f}\!=\!(f,0)$. Indeed, the stripe direction of model 2  is horizontal (Fig. \ref{fig-6}(f)) and $\eta_D$ is positive (Fig. \ref{fig-9}(d)) under the tensile stress of horizontal direction ${\vec f}\!=\!(f,0)$. In other words, the response of model 2 on the non-deformed lattice with uniaxial stress ${\vec f}\!=\!(f,0)$ is the same as that on the deformed lattice in Fig. \ref{fig-14}(b) corresponding to $\xi\!<\!1$. This is considered to be the reason why model 2 provides the consistent result of stripe direction with experimental data. 

From these observations, we find that the small value region of $v_{ij}$ plays an important role in the model's response. The small value region in model 1  is $\theta\!\simeq\! \pi/2$ (Figs. \ref{fig-19}(a),(b)), where $\tau_i$ is almost vertical to ${\vec{e}}_{ij}^{\;\prime}$, and $v_{ij}$ for  $\xi\!\not=\!1$ is almost the same as $v_{ij}$ for $\xi\!=\!1$, and therefore no new result is expected in model 1.  In contrast, the small value region in model 2 is  $\theta\!\simeq\! 0$ (Figs. \ref{fig-19}(c),(d)), where $\tau_i$ is almost parallel to ${\vec{e}}_{ij}^{\;\prime}$, and even a small deviation of $v_{ij}$ (dashed line) from $v_{ij}$ (solid line) is relevant. Such a non-trivial behavior of model 2 emerging from small $v_{ij}$ region is understood from the fact that the effective coupling constant $\lambda_{ij}$ is given by a rational function of $v_{ij}$.

We should emphasize that the result, supporting that model 2 is consistent
with both skyrmion deformation and stripe direction, is obtained by
comparing model 1 and model 2, and that the result of model 2 is consistent
with that in Ref. \cite{JWang-etal-PRB2018}, where an additional
energy term for MEC is included in a Landau-Ginzburg free energy.
In this additional interaction term, strains and magnetization are
directly coupled. In our models, the strain field $\tau$ is introduced
in $S_{f}$, and $\tau$ represents strain direction, though $S_{f}$ includes
no direct interaction of $\tau$ and magnetization or spin variable
$\sigma$. Thus, we consider that model 2 supports the model 
in Ref. \cite{Shibata-etal-Natnanotech2015},
where an anisotropy in the DMI coefficient is explicitly assumed,
implying that uniaxial stress deforms DMI anisotropic.

Another choice is that both FMI and DMI are modified by FG modeling prescription.
This model is certainly expected to reproduce the experimentally observed shape deformation of skyrmions. However, 
this choice is not suitable for reproducing the stripe direction alignment by lattice deformation because
model 1 is contradictory for this purpose, as demonstrated above. Therefore, we eliminate 
this choice from suitable models and find the conclusion stated above.

%----------------------------------------------------------
\section{Summary and conclusion}
\label{conclusion} 
%----------------------------------------------------------
Using a Finsler geometry (FG) model on a 2D triangular lattice with
periodic boundary conditions, we numerically study skyrmion deformation
under uniaxial stress and the lattice deformation.
 Two different
models, model 1 and model 2, are examined: the ferromagnetic energy
$S_{\rm FM}$ and Dzyaloshinskii-Moriya energy $S_{\rm DM}$ are
deformed by FG modeling prescription in model 1 and model 2, respectively.
In these FG models, the coupling constants $\lambda$ and $D$ of
$S_{\rm FM}$ and $S_{\rm DM}$ are dynamically deformed to be
direction-dependent such that $\lambda_{x}\!\not=\!\lambda_{y}$ and
$D_{x}\!\not=\!D_{y}$.
 In both models, the ratio $\lambda/D$ is dynamically distorted to be 
 direction dependent with a newly introduced internal degree of freedom $\tau$
for strains and a mechanical force or stress $\vec{f}$.

We find that the results of both models for skyrmion deformation under
uniaxial stress are consistent with the reported experimental data.
For the direction of stripes as a response to the stresses, the numerical
data of both models are also consistent with the reported experimental
result observed at room temperature with zero magnetic field. However,
we show that the responses of the two models to lattice deformations
are different from each other in the stripe phase. In this case, only
the data obtained by model 2 are shown to be consistent with the experimental
result. We conclude that in real systems only lattice deformations
due to the DMI are relevant. Note that the original model, in which
both FMI and DMI energies are not deformed by FG modeling prescription,
is also examined under the lattice deformations, and the produced
stripe directions are found to be different from those of the experimental
data. This shows that the lattice deformations naturally introduced
into the system by the FG modeling are necessary to explain the experimental
results.

Combining the obtained results for responses to both uniaxial stresses
and lattice deformations, we conclude that the anisotropy of the DMI
coefficient is considered to be the origin of the experimentally observed and reported
skyrmion deformations by uniaxial mechanical stresses. 
Thus, the FG modeling can provide a successful model 
to describe modulated chiral magnetic excitations on thin films caused by the anistropy in
 the ratios $\lambda/D$.

%----------------------------------------------------------
\acknowledgements 
%----------------------------------------------------------
This study was initiated during a two-month stay of S. E. H. at Ibaraki
KOSEN in 2017, and this stay was financially supported in part by
Techno AP Co. Ltd., Genesis Co. Ltd., Kadowaki Sangyo Co. Ltd, and
also by JSPS KAKENHI Grant Number JP17K05149. The author H.K. acknowledges
V. Egorov for simulation tasks in the early stage of this work during
a four-month stay from 2019 to 2020 at Sendai KOSEN. The simulations
and data analyses were performed with S. Tamanoe, S. Sakurai, and
Y. Tanaka's assistance. This work is supported in part by JSPS Grant-in-Aid
for Scientific Research on Innovative Areas "Discrete Geometric Analysis
for Materials Design": Grant Number 20H04647.

\appendix

\section{Finsler geometry modeling of ferromagnetic and Dzyaloshinskii-Moriya
interactions \label{FG-model}}

In this Appendix \ref{FG-model}, we show detailed information on
how the discrete forms of $S_{{\rm FM}}$ and $S_{{\rm DM}}$ in Eqs.
(\ref{model-1}) and (\ref{model-2}) are obtained. 
To simplify descriptions, we focus on the models on non-deformed lattices in Eqs.
(\ref{model-1}) and (\ref{model-2}). Note that descriptions of models on deformed lattices in  Eqs. (\ref{model-1prime}) and (\ref{model-2prime}) remain unchanged except the definition of $v_{ij}$. 
Let us start with
the continuous form of $S_{{\rm FM}}$. Since the variable $\sigma(\in S^{2}:{\rm unit\;sphere})$
is defined on a two-dimensional surface, the continuous $S_{{\rm FM}}$
and $S_{{\rm DM}}$ are given by 
\begin{eqnarray}
\begin{split} & S_{{\rm FM}}=\frac{1}{2}\int\sqrt{g}d^{2}xg^{ab}\frac{\partial\sigma}{\partial x^{a}}\cdot\frac{\partial\sigma}{\partial x^{b}},\\
 & S_{{\rm DM}}=\int\sqrt{g}d^{2}xg^{ab}\frac{\partial{\vec{r}}}{\partial x^{a}}\cdot\sigma\times\frac{\partial\sigma}{\partial x^{b}},
\end{split}
\end{eqnarray}
where $g^{ab}$ is the inverse of the metric $g_{ab}$, and $g$ is
its determinant (see also Ref. \cite{Diep-Koibuchi-Frustrated2020}).
Note that the unit tangential vector ${\vec{e}}_{a}$ can be used
for $\partial{\vec{r}}/\partial x^{a}$, which is not always a unit
vector. Indeed, the difference between ${\vec{e}}_{a}$ and $\partial{\vec{r}}/\partial x^{a}$
is a constant multiplicative factor on the regular triangular lattice,
and therefore, we use ${\vec{e}}_{a}$ for $\partial{\vec{r}}/\partial x^{a}$
for simplicity. For simulations on deformed lattices, this unit vector
${\vec{e}}_{a}$ is replaced by a more general one ${\vec{e}}_{a}^{\;\prime}$
in Eq. (\ref{new-bond-vect}).

%f-a1
\begin{figure}[h]
\centering{}\includegraphics[width=10.5cm]{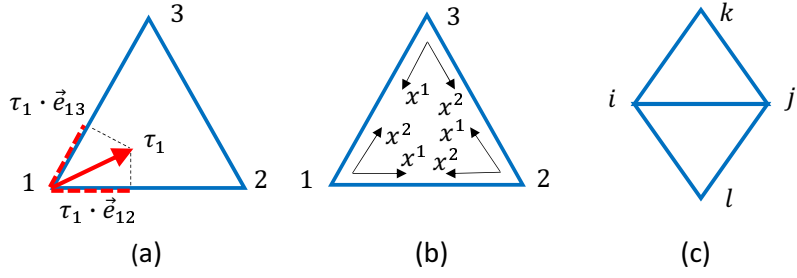}
\caption{(a) A triangle of vertices 123 and a strain field $\tau_{1}$ at vertex
1, and its tangential components $\tau_{1}\cdot{\vec{e}}_{12}$ and
$\tau_{1}\cdot{\vec{e}}_{13}$ along the directions ${\vec{e}}_{12}$
and ${\vec{e}}_{13}$, which are the unit tangential vectors from
vertices 1 to 2 and 1 to 3. (b) Three possible local coordinates on
the triangle 123, (c) two neighboring triangles $ijk$ and $jil$. }
\label{fig-A-1} 
\end{figure}

Here we assume that $g_{ab}$ is not always limited to the induced
metric $(\partial{\vec{r}}/\partial x^{i})\cdot(\partial{\vec{r}}/\partial x^{j})$,
but it is assumed to be of the form 
\begin{eqnarray}
g_{ab}=\begin{pmatrix}v_{12}^{-2} & 0\\
0 & v_{13}^{-2}
\end{pmatrix}\label{Finsler-metric}
\end{eqnarray}
on the triangle of vertices 123 (see Fig. \ref{fig-A-1}(a)), where
$v_{ij}$ is defined by using the strain field $\tau_{i}(\in S^{1}:{\rm unit\;circle})$
such that 
\begin{eqnarray}
\begin{split} & v_{ij}=|\tau_{i}\cdot{\vec{e}}_{ij}|+v_{0},\quad({\rm for}\;S_{{\rm FM}};\;{\rm model\;1}),\\
 & v_{ij}=\sqrt{1-\left(\tau_{i}\cdot{\vec{e}}_{ij}\right)^{2}}+v_{0},\quad({\rm for}\;S_{{\rm DM}};\;{\rm model\;2}).
\end{split}
\label{Finsler-velocity}
\end{eqnarray}
Note that the definition of $v_{ij}$ in $S_{{\rm FM}}$ in model
1 is different from that in $S_{{\rm DM}}$ in model 2. 

We should comment that the usage of  Finsler geometry in this paper for chiral magnetism is not the standard one of non-Euclidean geometry such as in Ref. \cite{Gaididei-etal-PRL2014}. In the case of Ref. \cite{Gaididei-etal-PRL2014}, a non-flat geometry is assumed to describe real curved thin films in ${\bf R}^3$ and to extract curvature effect on a magnetic system. In contrast, the film in this paper is flat and follows Euclidean geometry; however, an additional distance called Finsler length is introduced to describe Hamiltonian $S_{{\rm FM}}$  or $S_{{\rm DM}}$.  Even when the surface is curved, in which the surface geometry follows the induced metric or Euclidean geometry in ${\bf R}^3$ as in Ref. \cite{Gaididei-etal-PRL2014}, a Finsler length can also be introduced in addition to the surface geometry.  Such a non-Euclidean length scale can constantly be introduced to the tangential space, where the length of vector or the distance of two different points is defined by the newly introduced metric tensor such as $g_{ab}$ in Eq. (\ref{Finsler-metric}).  Therefore, in the FG modeling prescription, we have two different length scales; one is the Euclidean length for thin films in ${\bf R}^3$ and the other is dynamically changeable Finsler length for Hamiltonian. 

The Finsler length scale is used to effectively deform the coefficient $\lambda_{ij}$ in Eqs. (\ref{model-1}), (\ref{model-2}), which will be described below in detail. This $\lambda_{ij}$ varies depending on the internal strain variable $\tau$, which is integrated out in the partition function, and therefore, all physical quantities are effectively integrated over different length scales characterized by the ratio $\lambda/D$ of interaction coefficients for FMI and DMI. Here, this ratio is fluctuating and its mean value can be observed and expressed by using the effective coupling constant in Eq. (\ref{anisotropy-effective-D}).  Thus, ``dynamically deformed $D$'' means that all-important length scales are effectively integrated out with the Boltzmann weight to calculate observable quantities. Note that this is possible if $g_{ab}$ is treated to be dynamically changeable. 
For this reason, this FG modeling is effective, especially for anisotropic phenomena, because we can start with isotropic models such as the isotropic FMI and DMI. Therefore, the FG model is in sharp contrast to those models with explicit anisotropic interaction terms such as Landau-type theory for MEC. This FG modeling is coarse-grained one like the linear chain model, of which the connection to monomers is mathematically confirmed \cite{Doi-Edwards-1986}.   In such a coarse-grained modeling, the detailed information on electrons and atoms are lost from the beginning like in the case of FMI. In other words,  no specific information at the scale of atomic level is necessary to calculate physical quantities even in such complex anisotropic phenomena.  

To obtain the discrete expressions of $S_{{\rm FM}}$, we replace
$\int\sqrt{g}d^{2}x\to\sum_{\Delta}(1/v_{12}v_{13})$ and $g^{11}\partial\sigma/\partial x^{1}\cdot\partial\sigma/\partial x^{1}\to v_{12}^{2}(\sigma_{2}-\sigma_{1})^{2}$,
$g^{22}\partial\sigma/\partial x^{2}\cdot\sigma/\partial x^{2}\to v_{13}^{2}(\sigma_{3}-\sigma_{1})^{2}$
on the triangle of vertices 123 (Fig. \ref{fig-A-1}(a)), where the
local coordinate origin is at vertex 1, and $\sum_{\Delta}$ denotes
the sum over triangles. The discrete form of $S_{{\rm DM}}$ is also
obtained by the replacements $g^{11}\partial{\vec{r}}/\partial x^{1}\cdot(\sigma\times{\partial\sigma}/{\partial x^{1}})\to v_{12}^{2}{\bf e}_{12}\cdot(\sigma_{1}\times\sigma_{2})$,
$g^{22}\partial{\vec{r}}/\partial x^{2}\cdot(\sigma\times{\partial\sigma}/{\partial x^{2}})\to v_{13}^{2}{\bf e}_{13}\cdot(\sigma_{1}\times\sigma_{3})$.
Then, we have 
\begin{eqnarray}
\begin{split}S_{{\rm FM}} & =\frac{1}{2}\int\sqrt{g}d^{2}x\left(g^{11}\frac{\partial\sigma}{\partial x^{1}}\cdot\frac{\partial\sigma}{\partial x^{1}}+g^{22}\frac{\partial\sigma}{\partial x^{2}}\cdot\frac{\partial\sigma}{\partial x^{2}}\right)\\
 & \to\sum_{\Delta}\left[\frac{v_{12}}{v_{13}}\left(1-\sigma_{1}\cdot\sigma_{2}\right)+\frac{v_{13}}{v_{12}}\left(1-\sigma_{1}\cdot\sigma_{3}\right)\right],
\end{split}
\label{discrete-SFM-1}
\end{eqnarray}
and 
\begin{eqnarray}
\begin{split}S_{{\rm DM}} & =\int\sqrt{g}d^{2}x\left(g^{11}\frac{\partial{\vec{r}}}{\partial x^{1}}\cdot\sigma\times\frac{\partial\sigma}{\partial x^{1}}+g^{22}\frac{\partial{\vec{r}}}{\partial x^{2}}\cdot\sigma\times\frac{\partial\sigma}{\partial x^{2}}\right)\\
 & \to\sum_{\Delta}\left[\frac{v_{12}}{v_{13}}\left({\vec{e}}_{12}\cdot\sigma_{1}\times\sigma_{2}\right)+\frac{v_{12}}{v_{13}}\left({\vec{e}}_{13}\cdot\sigma_{1}\times\sigma_{3}\right)\right].
\end{split}
\label{discrete-SDM-1}
\end{eqnarray}
The local coordinate origin can also be assumed at vertices 2 and
3 on the triangle 123 (Fig. \ref{fig-A-1}(b)). Therefore, summing
over the discrete expressions of $S_{{\rm FM}}$ and $S_{{\rm DM}}$
for the three possible local coordinates, which are obtained by replacing
the indexes $1\to2,2\to3,\cdots$ with the factor $1/3$, we have
\begin{eqnarray}
\begin{split}S_{{\rm FM}}=\frac{1}{3}\sum_{\Delta} & \left[\left(\frac{v_{12}}{v_{13}}+\frac{v_{21}}{v_{23}}\right)\left(1-\sigma_{1}\cdot\sigma_{2}\right)+\left(\frac{v_{23}}{v_{21}}+\frac{v_{32}}{v_{31}}\right)\left(1-\sigma_{2}\cdot\sigma_{3}\right)\right.\\
 & +\left.\left(\frac{v_{13}}{v_{12}}+\frac{v_{31}}{v_{32}}\right)\left(1-\sigma_{3}\cdot\sigma_{1}\right)\right],
\end{split}
\label{discrete-SFM-2}
\end{eqnarray}
and 
\begin{eqnarray}
\begin{split}S_{{\rm DM}}=\frac{1}{3}\sum_{\Delta} & \left[\left(\frac{v_{12}}{v_{13}}+\frac{v_{21}}{v_{23}}\right)\left({\vec{e}}_{12}\cdot\sigma_{1}\times\sigma_{2}\right)+\left(\frac{v_{23}}{v_{21}}+\frac{v_{32}}{v_{31}}\right)\left({\vec{e}}_{23}\cdot\sigma_{2}\times\sigma_{3}\right)\right.\\
 & +\left.\left(\frac{v_{13}}{v_{12}}+\frac{v_{31}}{v_{32}}\right)\left({\vec{e}}_{31}\cdot\sigma_{3}\times\sigma_{1}\right)\right].
\end{split}
\label{discrete-SDM-2}
\end{eqnarray}
Replacing the vertices 1,2,3 with $i,j,k$, we have the following
expressions for $S_{{\rm FM}}$ and $S_{{\rm DM}}$ such that 
\begin{eqnarray}
\begin{split} & S_{{\rm FM}}=\sum_{\Delta}\left[\lambda_{ij}\left(1-\sigma_{i}\cdot\sigma_{j}\right)+\lambda_{jk}\left(1-\sigma_{j}\cdot\sigma_{k}\right)+\lambda_{ki}\left(1-\sigma_{k}\cdot\sigma_{i}\right)\right],\\
 & S_{{\rm DM}}=\sum_{\Delta}\left[\lambda_{ij}\left({\vec{e}}_{ij}\cdot\sigma_{i}\times\sigma_{j}\right)+\lambda_{jk}\left({\vec{e}}_{jk}\cdot\sigma_{j}\times\sigma_{k}\right)+\lambda_{ki}\left({\vec{e}}_{ki}\cdot\sigma_{k}\times\sigma_{i}\right)\right],\\
 & \lambda_{ij}=\frac{1}{3}\left(\frac{v_{ij}}{v_{ik}}+\frac{v_{ji}}{v_{jk}}\right),
\end{split}
\label{discrete-SFM-SDM}
\end{eqnarray}
where $k$ in $\lambda_{ij}$ is the third vertex number other than
$i$ and $j$. Note that $\lambda_{ij}\!=\!\lambda_{ji}$ is satisfied.

The sum over triangles $\sum_{\Delta}$ in these expressions can also
be replaced by the sum over bonds $\sum_{ij}$, and we also have 
\begin{eqnarray}
S_{{\rm FM}}=\sum_{ij}\bar{\lambda}_{ij}\left(1-\sigma_{i}\cdot\sigma_{j}\right),\quad S_{{\rm DM}}=\sum_{ij}\bar{\lambda}_{ij}\left({\bf e}_{ij}\cdot\sigma_{i}\times\sigma_{j}\right),\label{discrete-SFDM}
\end{eqnarray}
where the coefficients $\bar{\lambda}_{ij}$ on the triangles are
given by 
\begin{eqnarray}
\bar{\lambda}_{ij}=\frac{1}{3}\left(\frac{v_{ij}}{v_{ik}}+\frac{v_{ji}}{v_{jk}}+\frac{v_{ij}}{v_{il}}+\frac{v_{ji}}{v_{jl}}\right).
\end{eqnarray}
In this expression, the vertices $k$ and $l$ are those connected
with $i$ and $j$ (see Fig. \ref{fig-A-1}(c)). The coefficient $\bar{\lambda}_{ij}$
is also symmetric; $\bar{\lambda}_{ij}\!=\!\bar{\lambda}_{ji}$, where
$k$ and $l$ should also be replaced by each other if $i$ is replaced
by $j$. For numerical implementation, the expressions in the sum
of triangles are easier than the sum over bonds, and we use the sum
over triangles in the simulations in this paper.

%f-a2
\begin{figure}[h]
\centering{}\includegraphics[width=8.5cm]{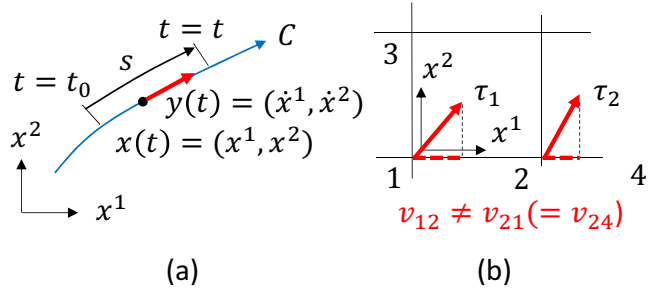}
\caption{(a) A curve $C$ parameterized by $t$ on a two-dimensional continuous
surface, where a point $x(t)=(x^{1},x^{2})$ on $C$ and its derivative
$y(t)=(\dot{x}^{1},\dot{x}^{2})$ are represented by a local coordinate.
(b) A regular square lattice with a local coordinate axes $x^{1}$
and $x^{2}$ at vertex 1 and strain fields $\tau_{i}(i\!=\!1,2)$
at vertices 1 and 2. Note that $v_{12}\!\protect\not=\!v_{21}$ implying
that the velocity from 1 to 2 is different from the velocity from
2 to 1, while $v_{21}\!=\!v_{24}$. }
\label{fig-A-2} 
\end{figure}

Now, the origin of the form of $g_{ab}$ in Eq. (\ref{Finsler-metric})
is briefly explained \cite{SS-Chern-AMS1996,Matsumoto-SKB1975,Bao-Chern-Shen-GTM200,Koibuchi-PhysA2014}.
Let $L(x(t),y(t))$ be a Finsler function on a two-dimensional surface
defined by 
\begin{eqnarray}
\begin{split} & L(x(t),y(t))=\sqrt{(y^{1})^{2}+(y^{2})^{2}}/|{\vec{v}}|=\sqrt{\left(\frac{dx^{1}}{dt}\right)^{2}+\left(\frac{dx^{2}}{dt}\right)^{2}}/|{\vec{v}}|,\\
 & |{\vec{v}}|=\sqrt{\left(\frac{dx^{1}}{ds}\right)^{2}+\left(\frac{dx^{2}}{ds}\right)^{2}},\quad{\vec{v}}=\left(\frac{dx^{1}}{ds},\frac{dx^{2}}{ds}\right),
\end{split}
\label{Finsler-function}
\end{eqnarray}
where ${\vec{v}}$ is a velocity along $C$ other than $y(t)\!=\!\left({dx^{1}}/{dt},{dx^{2}}/{dt}\right)$,
and ${\vec{v}}$ is assumed to be identical to the derivative of $(x^{1},x^{2})$
with respect to the parameter $s$ (Fig. \ref{fig-A-2}(a)). It is
easy to check that 
\begin{eqnarray}
s=\int_{t_{0}}^{t}L(x(t),y(t))dt\quad\left(\Leftrightarrow\frac{ds}{dt}=L(x(t),y(t))\right),\label{Finsler-length}
\end{eqnarray}
and this $s$ is called \textit{Finsler length} along the positive
direction of $C$. The Finsler metric $g_{ab},(a,b=1,2)$, which is
a $2\times2$ matrix, is given by using the Finsler function such
that 
\begin{eqnarray}
g_{ab}=\frac{1}{2}\frac{\partial^{2}L}{\partial y^{a}\partial y^{b}}.\label{metric-from-L}
\end{eqnarray}

Now, let us consider the Finsler function $L(x,y)$ on the square
lattice (for simplicity). Note that $L$ is defined only on the local
coordinate axes on the lattice, and therefore we have 
\begin{eqnarray}
L(x(t),y(t))=y^{1}/v_{12}\label{Finsler-f-on-x1}
\end{eqnarray}
on $x^{1}$ axis from vertices 1 to 2 (Fig. \ref{fig-A-2}(b)), where
$v_{12}$ is the velocity from vertex 1 to vertex 2 defined in Eq.
(\ref{Finsler-velocity}). From this expression and Eq. (\ref{metric-from-L}),
we have $g_{11}\!=\!v_{12}^{-2}$. We also have $g_{22}\!=\!v_{13}^{-2}$
from the Finsler function $L\!=\!y^{2}/v_{13}$ defined on $x^{2}$
axis from vertex 1 to vertex 3. Thus, we have the discrete and local
coordinate expression of Finsler metric in Eq. (\ref{Finsler-metric})
on square lattices shown in Fig. \ref{fig-A-2}(b), though
the expression of $g_{ab}$ in Eq. (\ref{Finsler-metric}) for triangular
lattices. Indeed, on triangular lattices, the expression of $g_{ab}$
is the same as that on square lattices, and the only difference is
that there are three possible local coordinates on triangles, while
there are four possible local coordinates on squares. Due to this
difference, the coefficient $\lambda_{ij}$ in Eq. (\ref{discrete-SFM-SDM})
becomes slightly different from that on square lattices; however,
we have no difference in the expression of $g_{ab}$ for the dependence
on the lattice structure.

\section{Graphical measurement of skyrmion shape anisotropy \label{skx-graphical}}

%f-b1
\begin{figure}[h]
\centering{}\includegraphics[width=10.5cm]{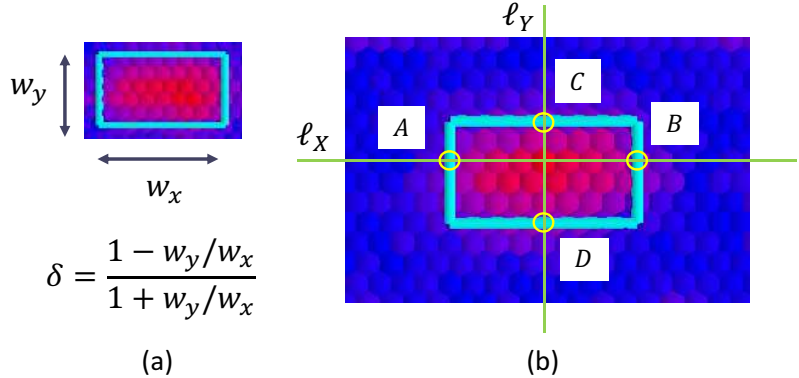}
\caption{(a) The definition of shape anisotropy $\delta$ with a snapshot of
skyrmion enclosed by a rectangle for the graphical measurement of
$w_{x}$ and $w_{y}$, and (b) two lines $\ell_{X}$ and $\ell_{Y}$,
passing through the local minimum of $\sigma_{z}$, are used to find
the four points $A$, $B$, $C$ and $D$ for the rectangle. }
\label{fig-B-1} 
\end{figure}

Here we describe how to measure the side lengths $w_{x}$ and $w_{y}$
of a skyrmion for the shape anisotropy $\delta$ (Fig.\ref{fig-B-1}(a)),
where a snapshot of the skyrmion is shown simply by two-color gradation
in blue and red using $\sigma_{z}(\in[-1,1])$. Two lines $\ell_{X}$
and $\ell_{Y}$ in Fig. \ref{fig-B-1}(b) are drawn parallel to the
$x$ and $y$ directions, and the point where two lines  cross is
a vertex where $\sigma_{z}$ is the local minimum (or maximum depending
on the direction of $\vec{B}$). This local minimum $\sigma_{z}$
is numerically determined to be smaller than those of the four nearest
neighbor vertices in all directions. The point $A$ is the first vertex,
where the sign of $\sigma_{z}$ changes from minus to plus, encountered
moving along $\ell_{X}$ from the crossing point. The other vertex
$B$ is also uniquely determined in the same way. Note that the crossing
point is not always located at the center of $A$ and $B$ on $\ell_{X}$.
The vertices $C$ and $D$ on $\ell_{Y}$ are also uniquely determined.
The values of $\sigma_{z}$ at these four points $A$, $B$, $C$
and $D$ are not always exactly identical to $\sigma_{z}\!=\!0$ but
are small positive close to $\sigma_{z}\!=\!0$.

\section*{References}

%-----------------------------------------------

\end{document}